\documentclass[ twocolumn, superscriptaddress, amsmath,amssymb, prc, aps]{revtex4-1}

\usepackage{amsmath}
\usepackage{romanbar}
\usepackage{graphicx}% Include figure files
\usepackage{dcolumn}% Align table columns on decimal point
\usepackage{xcolor}% color
\usepackage{bm}% bold math
\usepackage{SIunits}
\usepackage{array}
\usepackage{hyperref} 
\begin{document}

\title{Identification of new isomers in $^{228}$Ac : Impact on dark matter searches}

\author{K.~W.~Kim}
\affiliation{Center for Underground Physics, Institute for Basic Science (IBS), Daejeon 34126, Republic of Korea}
\author{G.~Adhikari}
\affiliation{Department of Physics, University of California San Diego, La Jolla, CA 92093, USA}
\author{E.~Barbosa~de~Souza}
\affiliation{Department of Physics and Wright Laboratory, Yale University, New Haven, CT 06520, USA}
\author{N.~Carlin}
\affiliation{Physics Institute, University of S\~{a}o Paulo, 05508-090, S\~{a}o Paulo, Brazil}
\author{J.~J.~Choi}
\affiliation{Department of Physics and Astronomy, Seoul National University, Seoul 08826, Republic of Korea} 
\author{S.~Choi}
\affiliation{Department of Physics and Astronomy, Seoul National University, Seoul 08826, Republic of Korea} 
\author{M.~Djamal}
\affiliation{Department of Physics, Bandung Institute of Technology, Bandung 40132, Indonesia}
\author{A.~C.~Ezeribe}
\affiliation{Department of Physics and Astronomy, University of Sheffield, Sheffield S3 7RH, United Kingdom}
\author{L.~E.~Fran{\c c}a}
\affiliation{Physics Institute, University of S\~{a}o Paulo, 05508-090, S\~{a}o Paulo, Brazil}
\author{C.~Ha}
\affiliation{Department of Physics, Chung-Ang University, Seoul 06973, Republic of Korea}
\author{I.~S.~Hahn}
\affiliation{Department of Science Education, Ewha Womans University, Seoul 03760, Republic of Korea} 
\affiliation{Center for Exotic Nuclear Studies, Institute for Basic Science (IBS), Daejeon 34126, Republic of Korea}
\affiliation{IBS School, University of Science and Technology (UST), Daejeon 34113, Republic of Korea}
\author{E.~J.~Jeon}
\affiliation{Center for Underground Physics, Institute for Basic Science (IBS), Daejeon 34126, Republic of Korea}
\author{J.~H.~Jo}
\affiliation{Department of Physics and Wright Laboratory, Yale University, New Haven, CT 06520, USA}
\author{H.~W.~Joo}
\affiliation{Department of Physics and Astronomy, Seoul National University, Seoul 08826, Republic of Korea} 
\author{W.~G.~Kang}
\affiliation{Center for Underground Physics, Institute for Basic Science (IBS), Daejeon 34126, Republic of Korea}
\author{M.~Kauer}
\affiliation{Department of Physics and Wisconsin IceCube Particle Astrophysics Center, University of Wisconsin-Madison, Madison, WI 53706, USA}
\author{H.~Kim}
\affiliation{Center for Underground Physics, Institute for Basic Science (IBS), Daejeon 34126, Republic of Korea}
\author{H.~J.~Kim}
\affiliation{Department of Physics, Kyungpook National University, Daegu 41566, Republic of Korea}
\author{S.~H.~Kim}
\affiliation{Center for Underground Physics, Institute for Basic Science (IBS), Daejeon 34126, Republic of Korea}
\author{S.~K.~Kim}
\affiliation{Department of Physics and Astronomy, Seoul National University, Seoul 08826, Republic of Korea}
\author{W.~K.~Kim}
\affiliation{IBS School, University of Science and Technology (UST), Daejeon 34113, Republic of Korea}
\affiliation{Center for Underground Physics, Institute for Basic Science (IBS), Daejeon 34126, Republic of Korea}
\author{Y.~D.~Kim}
\affiliation{Center for Underground Physics, Institute for Basic Science (IBS), Daejeon 34126, Republic of Korea}
\affiliation{Department of Physics, Sejong University, Seoul 05006, Republic of Korea}
\affiliation{IBS School, University of Science and Technology (UST), Daejeon 34113, Republic of Korea}
\author{Y.~H.~Kim}
\affiliation{Center for Underground Physics, Institute for Basic Science (IBS), Daejeon 34126, Republic of Korea}
\affiliation{Korea Research Institute of Standards and Science, Daejeon 34113, Republic of Korea}
\affiliation{IBS School, University of Science and Technology (UST), Daejeon 34113, Republic of Korea}
\author{Y.~J.~Ko}
\email{yjko@ibs.re.kr}
\affiliation{Center for Underground Physics, Institute for Basic Science (IBS), Daejeon 34126, Republic of Korea}
\author{E.~K.~Lee}
\affiliation{Center for Underground Physics, Institute for Basic Science (IBS), Daejeon 34126, Republic of Korea}
\author{H.~Lee}
\affiliation{IBS School, University of Science and Technology (UST), Daejeon 34113, Republic of Korea}
\affiliation{Center for Underground Physics, Institute for Basic Science (IBS), Daejeon 34126, Republic of Korea}
\author{H.~S.~Lee}
\email{hyunsulee@ibs.re.kr}
\affiliation{Center for Underground Physics, Institute for Basic Science (IBS), Daejeon 34126, Republic of Korea}
\affiliation{IBS School, University of Science and Technology (UST), Daejeon 34113, Republic of Korea}
\author{H.~Y.~Lee}
\affiliation{Center for Underground Physics, Institute for Basic Science (IBS), Daejeon 34126, Republic of Korea}
\author{I.~S.~Lee}
\affiliation{Center for Underground Physics, Institute for Basic Science (IBS), Daejeon 34126, Republic of Korea}
\author{J.~Lee}
\affiliation{Center for Underground Physics, Institute for Basic Science (IBS), Daejeon 34126, Republic of Korea}
\author{J.~Y.~Lee}
\affiliation{Department of Physics, Kyungpook National University, Daegu 41566, Republic of Korea}
\author{M.~H.~Lee}
\affiliation{Center for Underground Physics, Institute for Basic Science (IBS), Daejeon 34126, Republic of Korea}
\affiliation{IBS School, University of Science and Technology (UST), Daejeon 34113, Republic of Korea}
\author{S.~H.~Lee}
\affiliation{IBS School, University of Science and Technology (UST), Daejeon 34113, Republic of Korea}
\affiliation{Center for Underground Physics, Institute for Basic Science (IBS), Daejeon 34126, Republic of Korea}
\author{S.~M.~Lee}
\affiliation{Department of Physics and Astronomy, Seoul National University, Seoul 08826, Republic of Korea} 
\author{D.~S.~Leonard}
\affiliation{Center for Underground Physics, Institute for Basic Science (IBS), Daejeon 34126, Republic of Korea}
\author{B.~B.~Manzato}
\affiliation{Physics Institute, University of S\~{a}o Paulo, 05508-090, S\~{a}o Paulo, Brazil}
\author{R.~H.~Maruyama}
\affiliation{Department of Physics and Wright Laboratory, Yale University, New Haven, CT 06520, USA}
\author{R.~J.~Neal}
\affiliation{Department of Physics and Astronomy, University of Sheffield, Sheffield S3 7RH, United Kingdom}
\author{S.~L.~Olsen}
\affiliation{Center for Underground Physics, Institute for Basic Science (IBS), Daejeon 34126, Republic of Korea}
\author{B.~J.~Park}
\affiliation{IBS School, University of Science and Technology (UST), Daejeon 34113, Republic of Korea}
\affiliation{Center for Underground Physics, Institute for Basic Science (IBS), Daejeon 34126, Republic of Korea}
\author{H.~K.~Park}
\affiliation{Department of Accelerator Science, Korea University, Sejong 30019, Republic of Korea}
\author{H.~S.~Park}
\affiliation{Korea Research Institute of Standards and Science, Daejeon 34113, Republic of Korea}
\author{K.~S.~Park}
\affiliation{Center for Underground Physics, Institute for Basic Science (IBS), Daejeon 34126, Republic of Korea}
\author{R.~L.~C.~Pitta}
\affiliation{Physics Institute, University of S\~{a}o Paulo, 05508-090, S\~{a}o Paulo, Brazil}
\author{H.~Prihtiadi}
\affiliation{Center for Underground Physics, Institute for Basic Science (IBS), Daejeon 34126, Republic of Korea}
\author{S.~J.~Ra}
\affiliation{Center for Underground Physics, Institute for Basic Science (IBS), Daejeon 34126, Republic of Korea}
\author{C.~Rott}
\affiliation{Department of Physics, Sungkyunkwan University, Suwon 16419, Republic of Korea}
\author{K.~A.~Shin}
\affiliation{Center for Underground Physics, Institute for Basic Science (IBS), Daejeon 34126, Republic of Korea}
\author{A.~Scarff}
\affiliation{Department of Physics and Astronomy, University of Sheffield, Sheffield S3 7RH, United Kingdom}
\author{N.~J.~C.~Spooner}
\affiliation{Department of Physics and Astronomy, University of Sheffield, Sheffield S3 7RH, United Kingdom}
\author{W.~G.~Thompson}
\affiliation{Department of Physics and Wright Laboratory, Yale University, New Haven, CT 06520, USA}
\author{L.~Yang}
\affiliation{Department of Physics, University of California San Diego, La Jolla, CA 92093, USA}
\author{G.~H.~Yu}
\affiliation{Department of Physics, Sungkyunkwan University, Suwon 16419, Republic of Korea}
\collaboration{COSINE-100 Collaboration}
\date{\today}

\begin{abstract}
		We report the identification of metastable isomeric states of $^{228}$Ac at 6.28\,keV, 6.67\,keV and 20.19\,keV,  
with lifetimes of an order of 100\,ns. 
These states are produced by the $\beta$-decay of $^{228}$Ra, a component of the $^{232}$Th decay chain, 
with $\beta$ Q-values of 39.52\,keV, 39.13\,keV and 25.61\,keV, respectively. 
Due to the low Q-value of $^{228}$Ra as well as the relative abundance of $^{232}$Th and their progeny in low background experiments, 
these observations potentially impact the low-energy background modeling of dark matter search experiments. 
\end{abstract}
\maketitle

\section{Introduction} 
Although numerous astronomical observations support the conclusion that most of the matter in the universe is invisible dark matter, an understanding of its nature and interactions remains elusive~\cite{Clowe:2006eq,Aghanim:2018eyx}.  The dark matter phenomenon might be attributable to new particles, such as weakly interacting massive particles~(WIMPs)~\cite{PhysRevLett.39.165,Goodman:1984dc}. 
Tremendous experimental efforts have been mounted to detect nuclei recoiling from WIMP-nucleus interactions, but no definitive signal has yet been observed~\cite{Undagoitia:2015gya,Schumann:2019eaa}.
This motivates searches for new types of dark matter~\cite{Essig:2011nj,Kim:2018veo,Chatterjee:2018mej,Ha:2018obm} that would produce different experimental signatures in detectors.

Typically, these dark matter signals can manifest themselves as an excess of event rates above a known background~\cite{Aprile:2020tmw,Adhikari:2018ljm,XMASS:2018bid}. 
Thus, these searches require a precise understanding of the background sources within a detector. 
Since typical dark matter models predict low energy signals below 10\,keV,  
a precise modeling of the background sources in the low-energy signal regions is crucial. 
For instance, the recently observed event excess seen by XENON1T~\cite{Aprile:2020tmw} can be 
explained not only by new physics interactions but also by tritium ($^{3}$H) contamination, which undergoes a low-energy $\beta$-decay (Q-value 18.6\,keV, half-life $t_{1/2}$ = 12.3\,years). 

\begin{figure}[!htb] 
\begin{center}
\includegraphics[width=0.9\columnwidth]{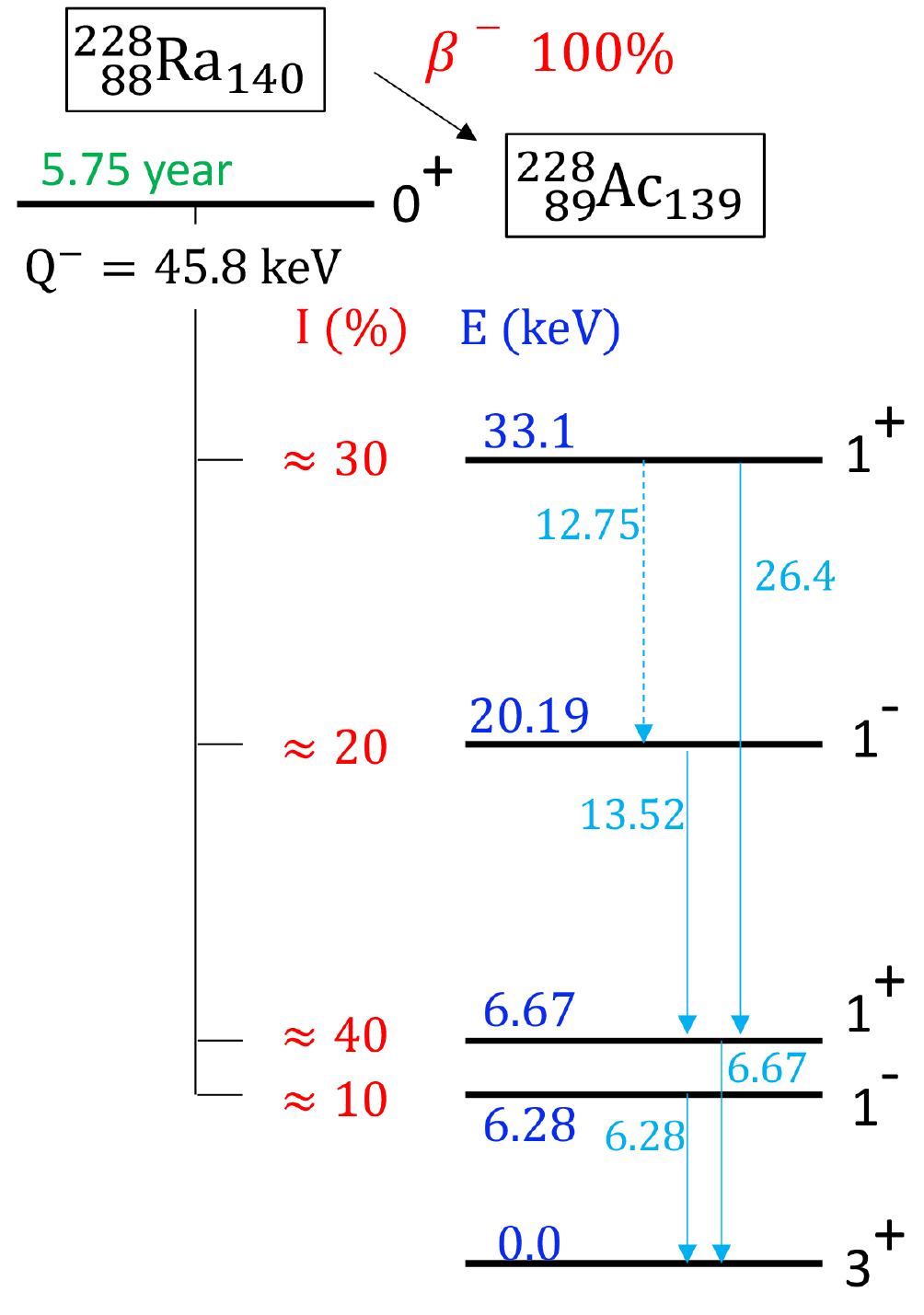}\\
\caption{ Nuclear level scheme for $^{228}$Ac adapted from Ref.~\cite{PhysRevC.52.88}.  
}
\label{fig_Ra228}
\end{center}
\end{figure}

Because of their long half-lives and large natural abundances, contamination from $^{238}$U and $^{232}$Th as well as their progenies are significant issues for  low-background dark matter search experiments~\cite{Lee:2007iq,Park:2020fsq}. 
Therefore, an accurate understanding of their contamination levels and resultant contributions to detector background is essential. 
Among these, $^{228}$Ra, which is produced by the $\alpha$-decay of $^{232}$Th, is of special interest due to the low total Q-value (45.8\,keV) of its decay. 
Based on data from the National Nuclear Data Center~\cite{nndc} and the Nuclear Data Sheet~\cite{Abusaleem:2014bsy}, $^{228}$Ra decays to $^{228}$Ac via $\beta$-particle emission~\cite{PhysRevC.52.88}. 
Since all $\beta$-decays from $^{228}$Ra are to excited states of $^{228}$Ac, the $\beta$-particle is always accompanied by a $\gamma$-ray or a conversion electron as described in the level scheme shown in Fig.~\ref{fig_Ra228}, which is based in part on measurements and a model-dependent analysis as discussed in \cite{PhysRevC.52.88}. 
Currently, the lifetimes of the excited states of $^{228}$Ac are unknown~\cite{nndc,Abusaleem:2014bsy,PhysRevC.52.88}. 
The odd numbers of protons and odd numbers of neutrons (odd-odd nuclei) suggest the possibility that these are multi-quasiparticle states and deformed nuclei that might result in isomeric states~\cite{Dracoulis:2016koy}. 
Even though the lifetime of these states have not been previously measured, Ref.~\cite{PhysRevC.52.88} pointed out that no coincident 6.67\,keV or 6.28\,keV $\gamma$ lines with 13.5\,keV or 26.4\,keV $\gamma$ lines were observed. 
This suggests the possible presence of long-lived isomeric states in $^{228}$Ac. 
If this is the case, the $\beta$ emission (e.g., Q = 39.52\,keV) and the following emission (e.g., Q = 6.28\,keV) will occur at different times and generate different signatures in the detector.
However, simulation programs commonly used by dark matter search experiments, such as Geant4~\cite{geant4}, model this decay with the simultaneous emission of the $\beta$ and the accompanying $\gamma$ or conversion electron. Failure to account for the isomeric
lifetime of the metastable state can have a significant impact on the background modeling and interpretation of dark matter search results. 

In this paper, we report the identification of the isomeric states in $^{228}$Ac using the COSINE-100 dark matter search detector~\cite{Adhikari:2017esn}. 
Due to a low contamination of $^{228}$Ra and about 250\,ns long scintillation decay time of the NaI(Tl) crystals, a quantitative evaluation of the lifetime, which is of order 100\,ns ($\mathcal{O}(100\,\mathrm{ns})$),  and the branching fraction for each state is difficult. 
However, we have studied how this might influence the understanding of the background in the low-energy signal region. 

\section{Experiment} 
COSINE-100~\cite{Adhikari:2017esn} consists of a 106\,kg array of eight ultra-pure NaI(Tl) crystals~\cite{cosinebg} each coupled to two photomultiplier tubes~(PMTs). 
The crystals are immersed in an active veto detector composed of 2,200\,L of a linear alkylbenzene~(LAB)-based liquid scintillator~(LS)~\cite{Park:2017jvs,Adhikari:2020asl}. 
These eight crystals are referred to as Crystal1 to Crystal8. 
Crystal1, Crystal5, and Crystal8 are excluded from this analysis due to their high background caused by high noise rate (Crystal1) and low light yield (Crystal5 and Crystal8)~\cite{Adhikari:2021szr}.
Data obtained between 20 October 2016 and 14 April 2020 are used for this analysis with a total exposure of 1181 live days. 

Signals from PMTs attached at each end of the crystals 
are digitized by 500\,MHz, 12-bit flash analog-to-digital converters. 
A trigger is generated when signals with amplitudes corresponding to one or more photoelectrons occur in both PMTs within a 200\,ns time window. 
The waveforms from the PMTs of all crystals are recorded when the trigger condition is satisfied by at least one crystal. 
The recorded waveform is 8\,$\mu$s long starting  2.4\,$\mu$s before occurrence of the trigger. Detailed descriptions of the COSINE-100 detector and its data acquisition system are provided elsewhere~\cite{Adhikari:2017esn,Adhikari:2018fpo}.

\section{Data Analysis}
\subsection{two-pulse events}
\begin{figure*}[!htb] 
\begin{center}
\begin{tabular}{ccc}
\includegraphics[width=0.33\textwidth]{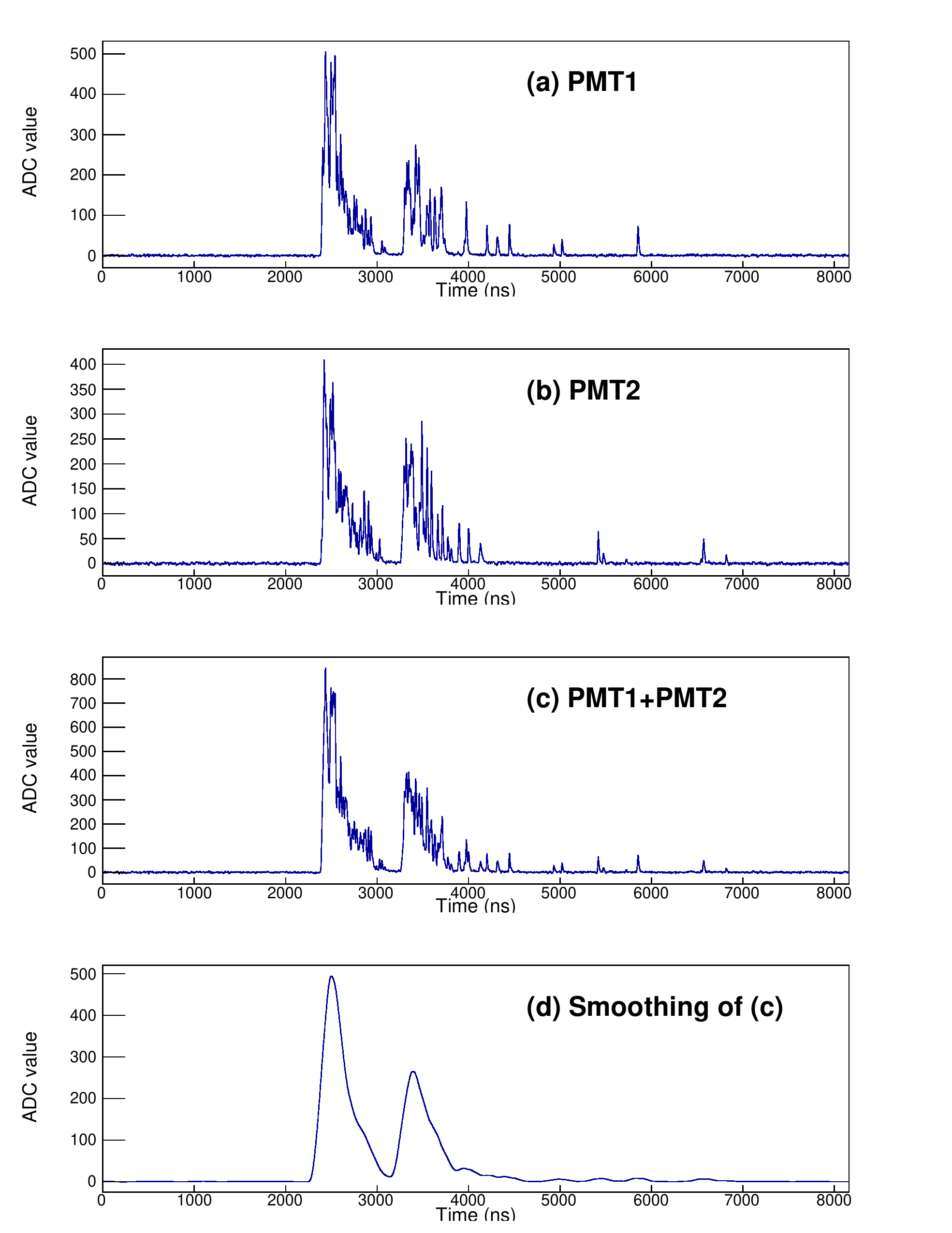}&
\includegraphics[width=0.33\textwidth]{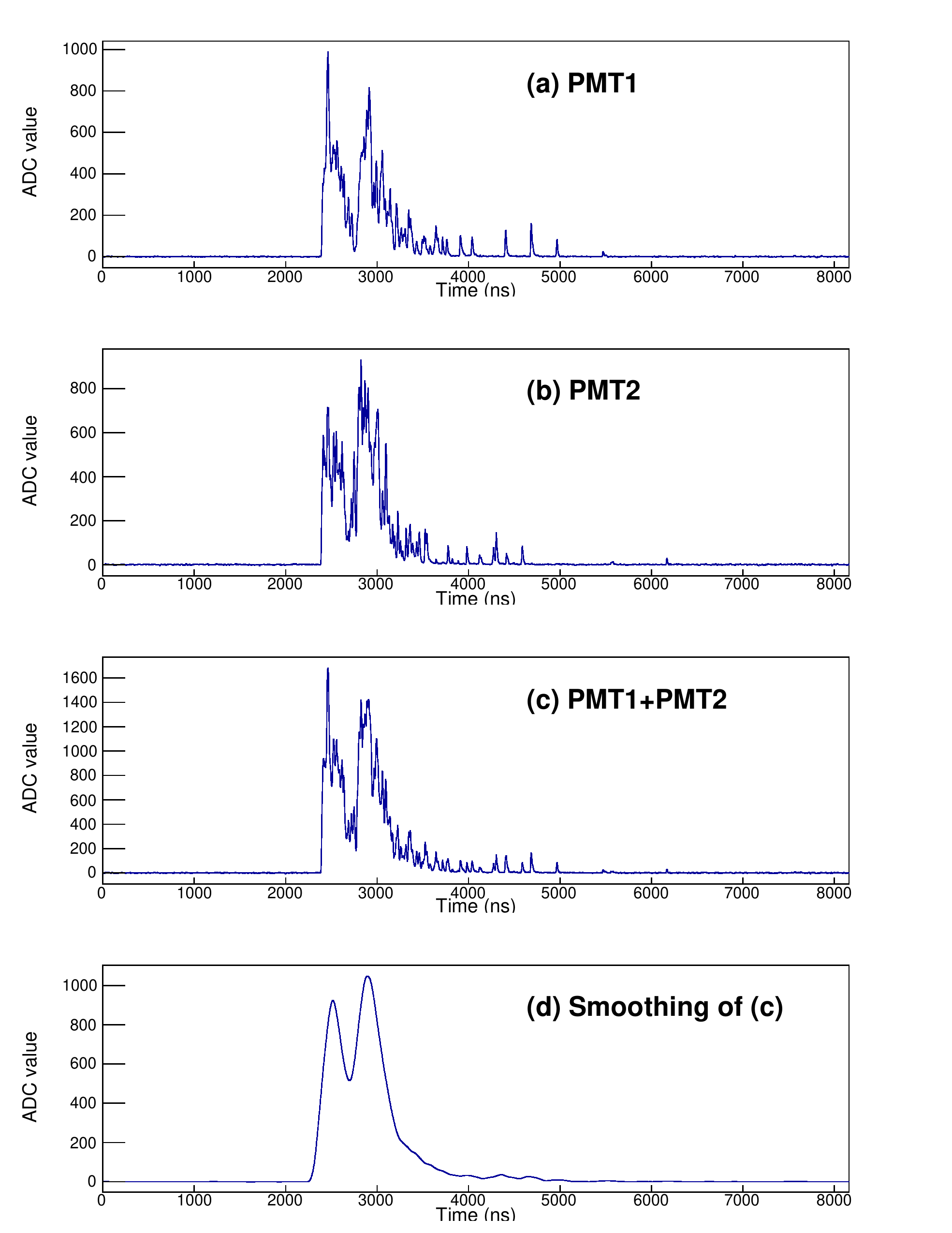} &
\includegraphics[width=0.33\textwidth]{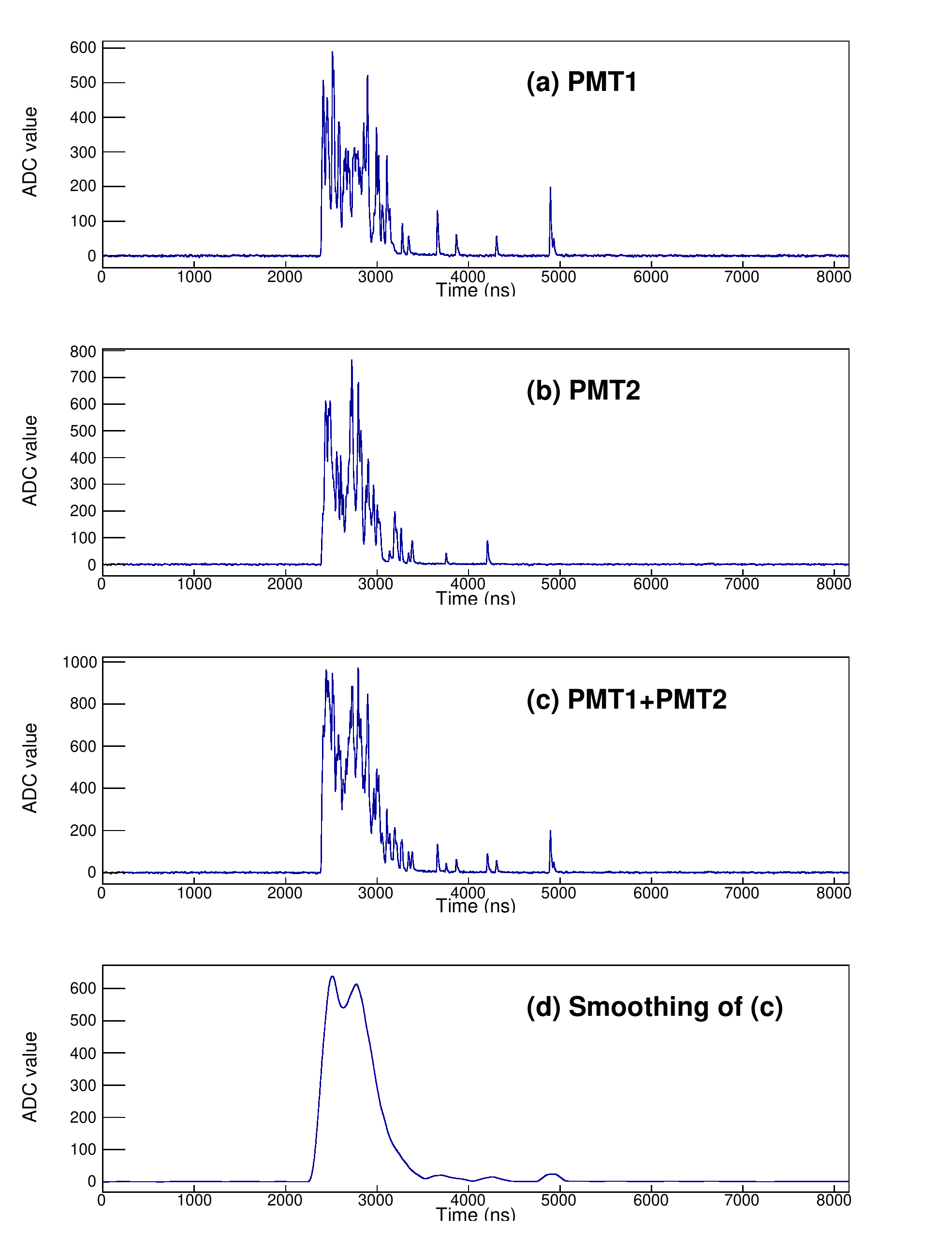} \\
Crystal 3 (Oct. 21, 2016) & Crystal 4 (Sept. 18, 2017)& Crystal 7 (Feb. 21, 2020)\\
\end{tabular}
\caption{
		Examples of selected two-pulse candidate events from different crystals on different dates. (a) and (b) are the signals from the individual PMTs, (c) displays the summed waveform of the two PMT signals,  and (d) shows the averaged 60\,ns bin  signals from the summed waveform. The offline selection of the two-pulse events is performed using the smoothed waveforms. 
}
\label{fig_twoevents}
\end{center}
\end{figure*}

In this analysis, we have studied selected events that contain two distinct pulses, named ``two-pulse events'', within the 8\,$\mu$s long waveforms. 
The event selection criteria to remove PMT-induced noise events~\cite{Adhikari:2020xxj} is applied as a preselection on candidate events.  
We have developed an offline selection algorithm to identify the two-pulse events using the summed waveforms from the two PMTs of each crystal. 
These  are smoothed by averaging 30 neighboring time bins into 60\,ns bin-width values. Two-pulse events are identified as those where, 
in addition to the initial rising edge of a pulse, there is a second rising edge of at least 1\,keV. 
The computed mean decay times in a 300\,ns time window starting from each rising edge are required to be greater than 100\,ns. 
In addition, we define the asymmetry between two PMT signals as $(Q_1 -Q_2)/(Q_1+Q_2)$ where $Q_1$ and $Q_2$ are charge measured by the two PMTs. 
We calculate the asymmetries for two identified pulses within 300\,ns time windows from the rising edges. 
The asymmetry for each pulse is required to be less than  0.25. A total of 4258 candidate events from the five crystals are accepted. 
Figure~\ref{fig_twoevents} shows three examples of the selected two-pulse candidate events that occurred in three different crystals.

\begin{figure*}[!htb] 
\begin{center}
\begin{tabular}{ccc}
\includegraphics[width=0.33\textwidth]{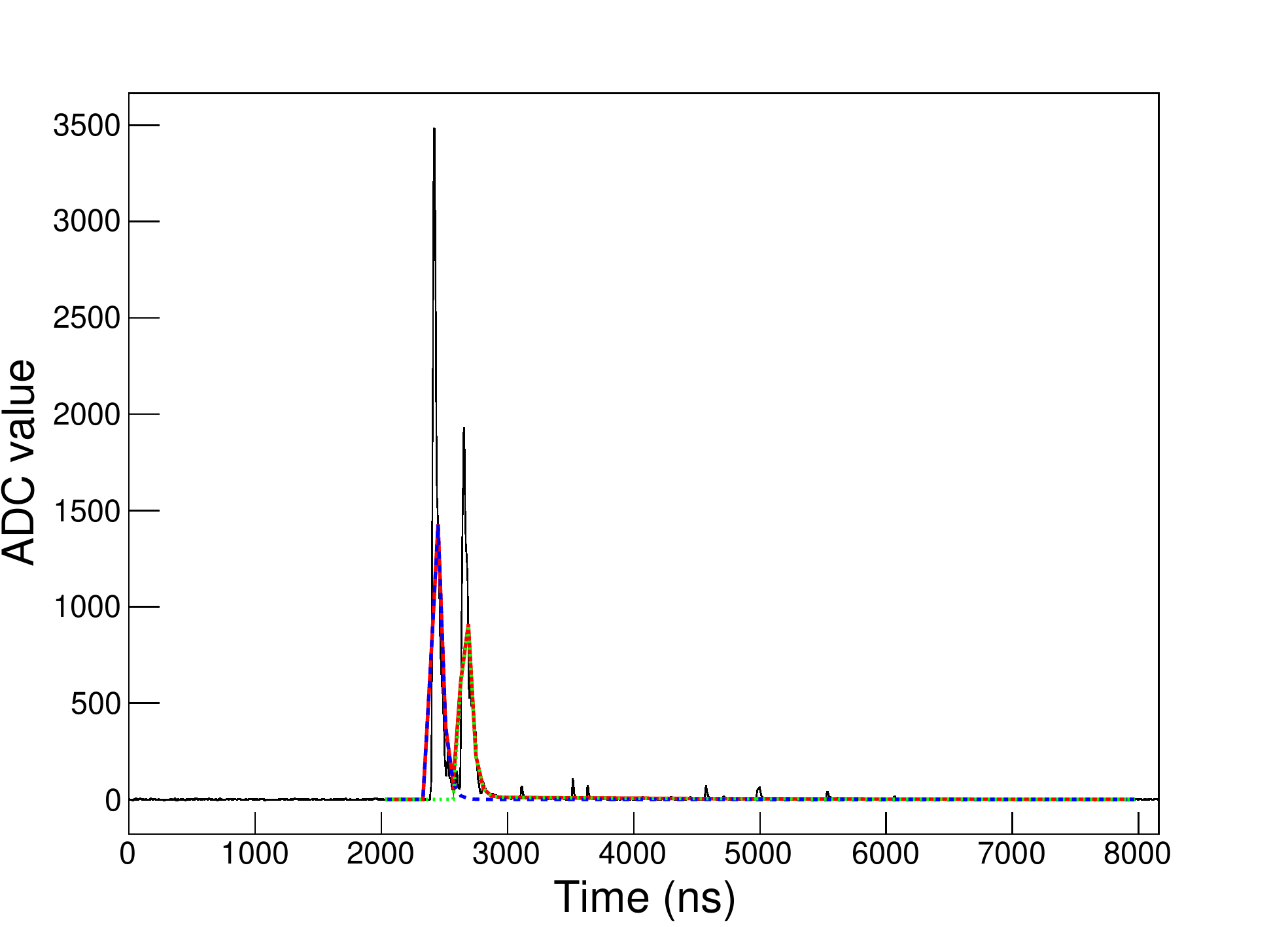}&
\includegraphics[width=0.33\textwidth]{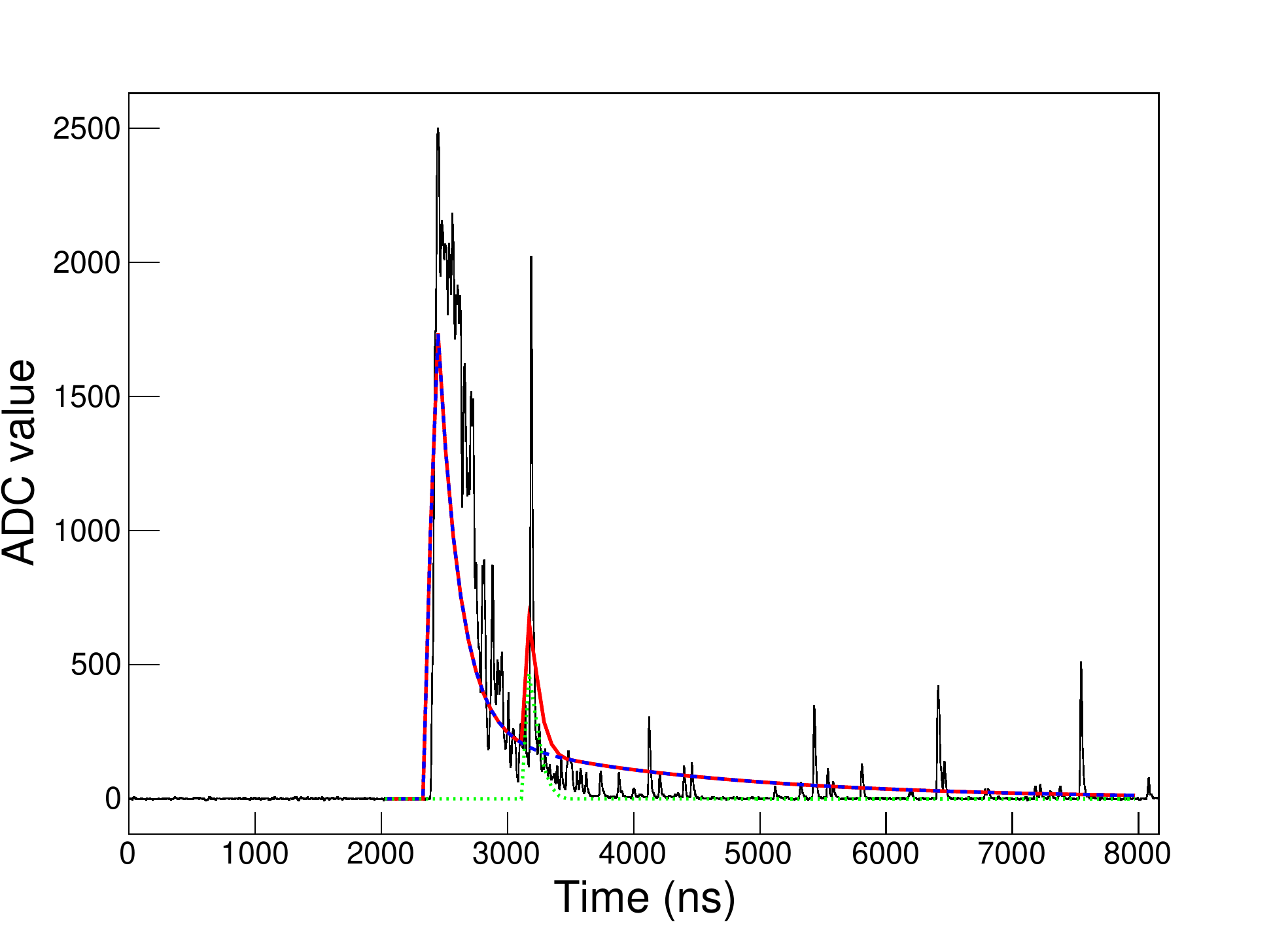} &
\includegraphics[width=0.33\textwidth]{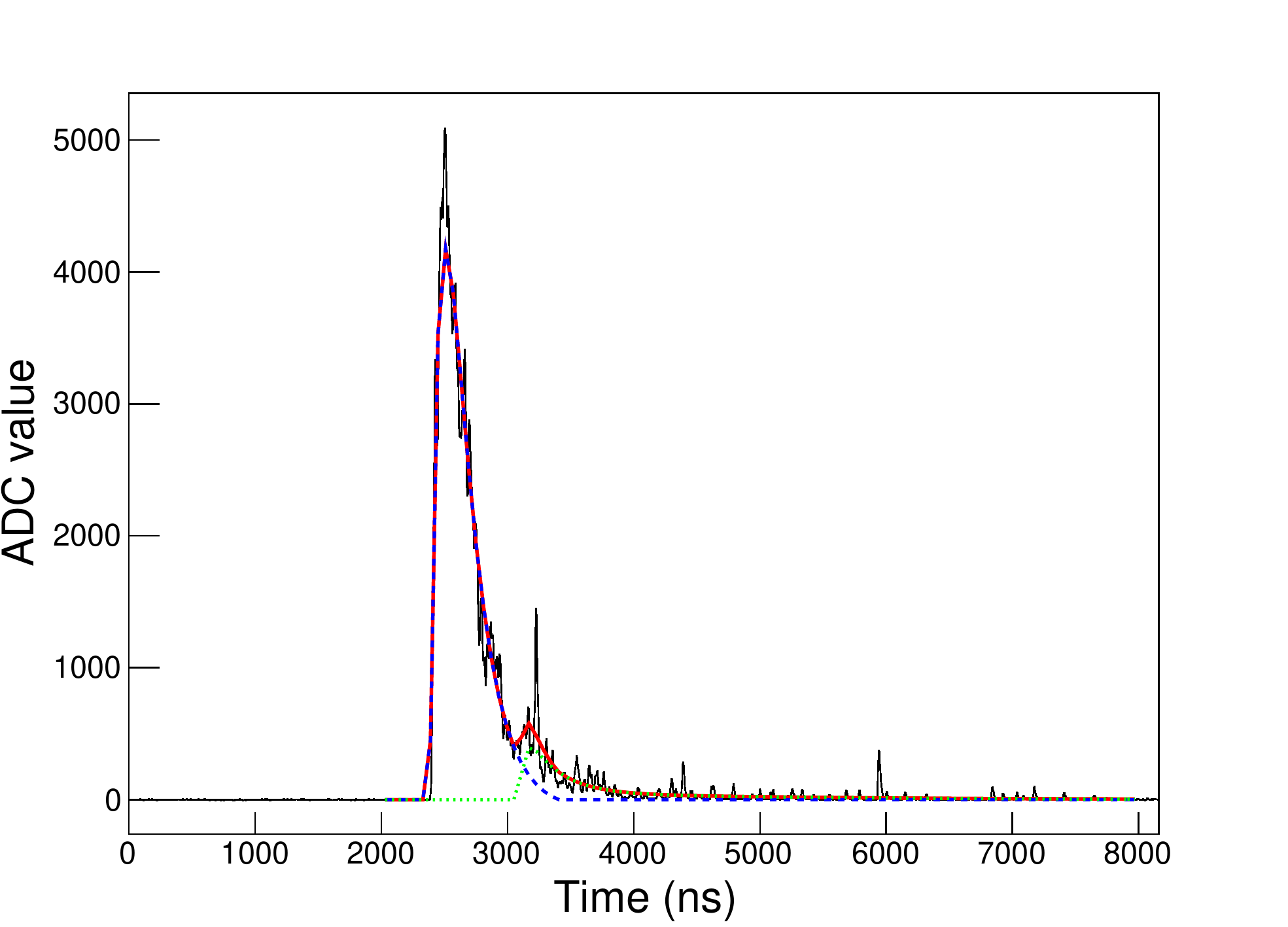} \\
\end{tabular}
\caption{
		Examples of two-pulse candidate events that are categorized as noisy events. 
}
\label{fig_badtwoevents}
\end{center}
\end{figure*}

Because of the low rate for two-pulse events, this initial selection still contains noisy events that did not originate from two scintillation occurrences. 
These include PMT-induced noise pulses or afterpulses that mimic  a second scintillation occurence and tail fluctuations due to reflections inside the crystal as shown in Fig.~\ref{fig_badtwoevents}. 
Further characterization of two-pulse events provides further discrimination of these noisy events. 

\begin{figure*}[!htb] 
\begin{center}
		\begin{tabular}{ccc}
\includegraphics[width=0.33\textwidth]{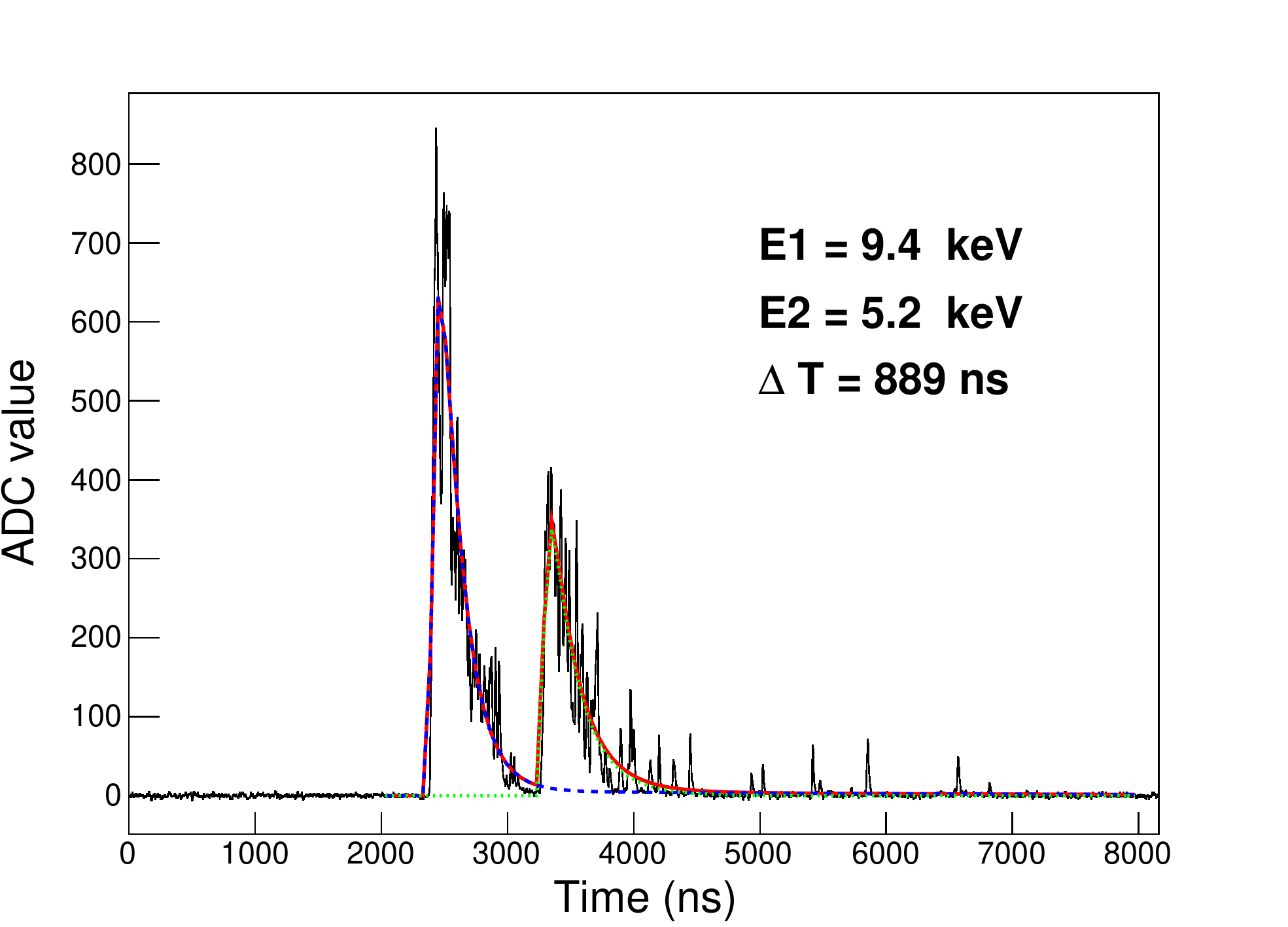} &
\includegraphics[width=0.33\textwidth]{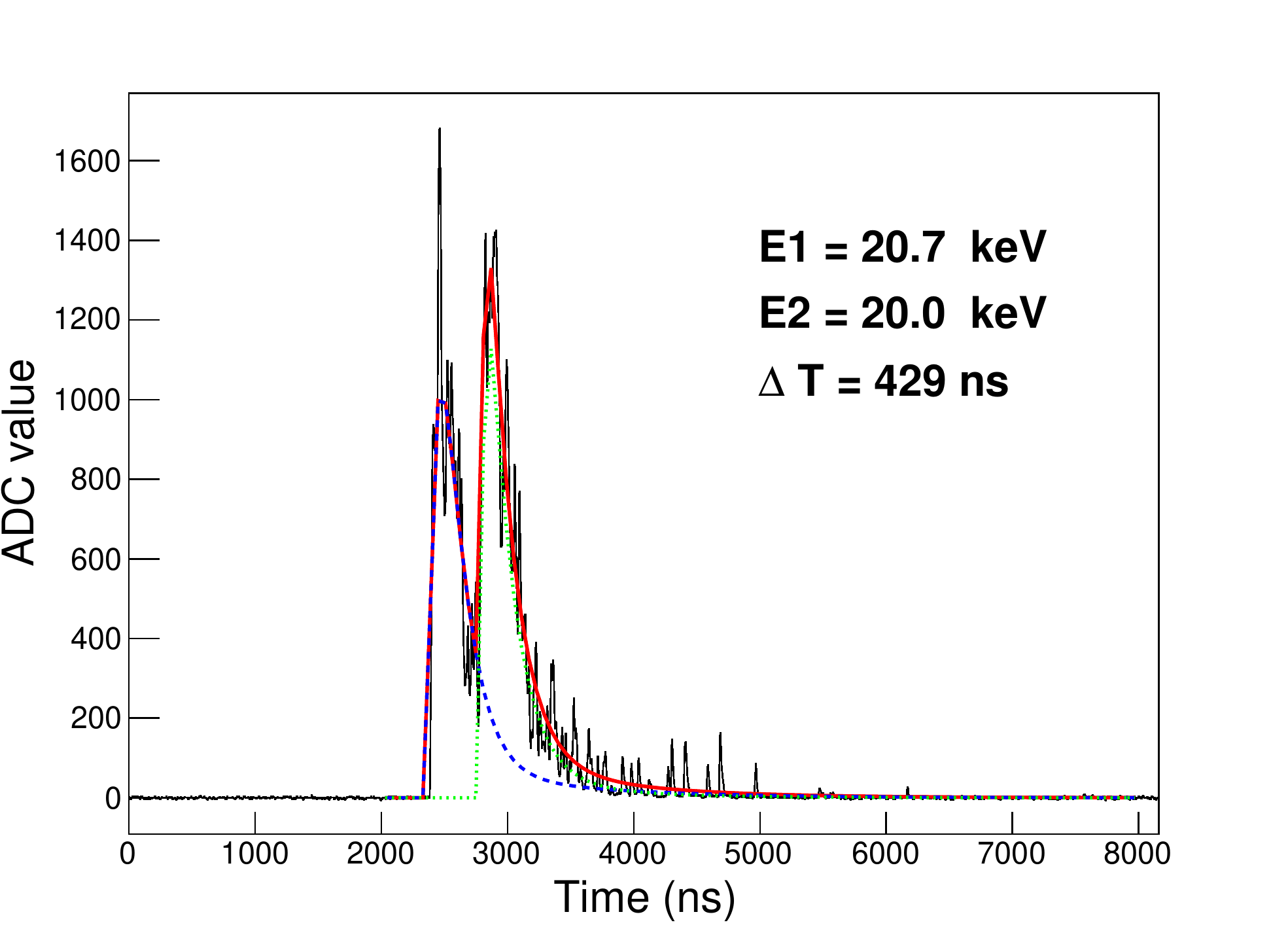} &
\includegraphics[width=0.33\textwidth]{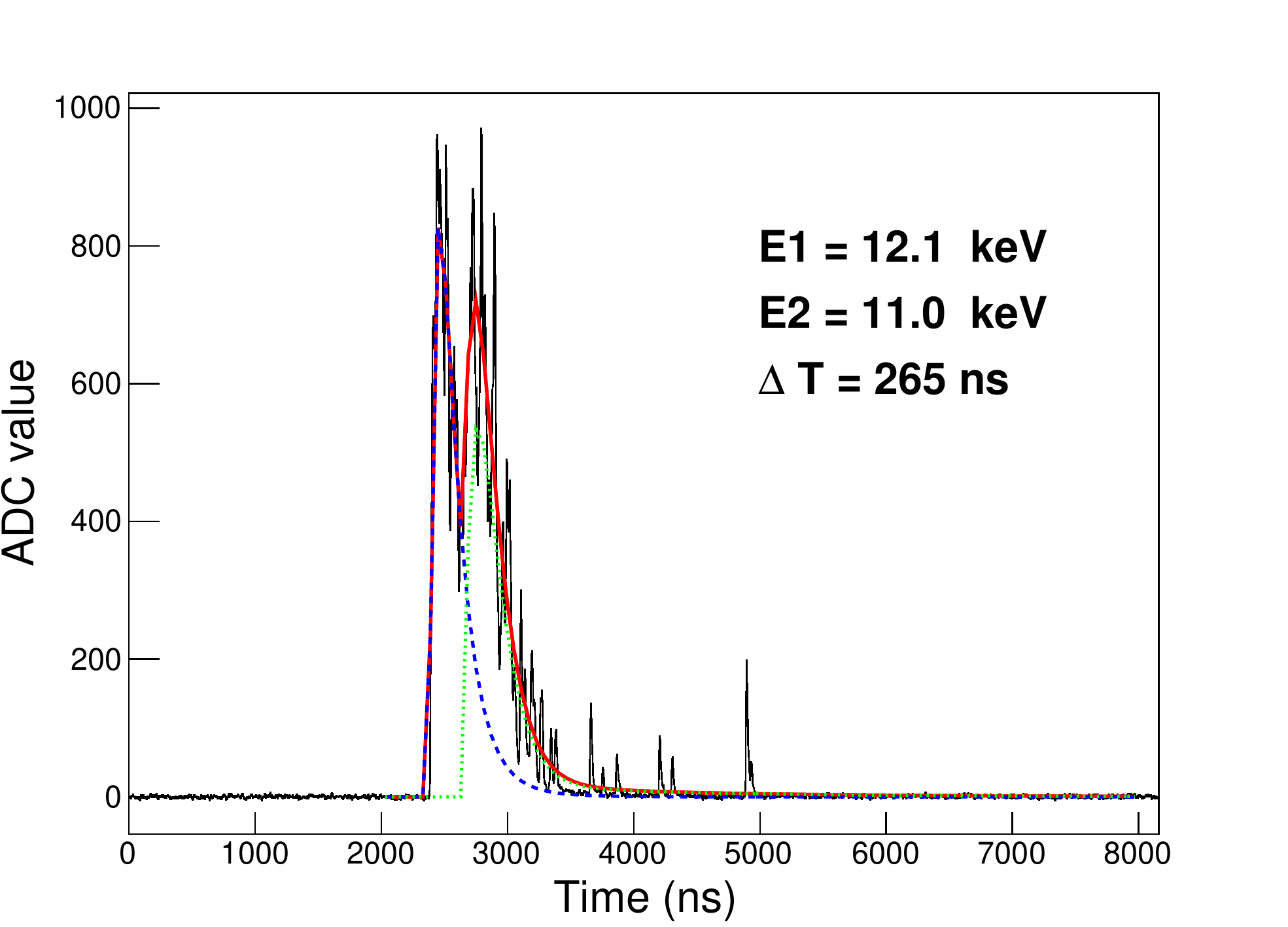}\\
Crystal 3 (Oct. 21, 2016) & Crystal 4 (Sept. 18, 2017)& Crystal 7 (Feb. 21, 2020)\\
\end{tabular}
\caption{
		The results of fits to the two pulses (lines) for the events in Fig.~\ref{fig_twoevents}. 
The energies of each pulse (E1 and E2) and the time difference ($\Delta$T) are listed. 
}
\label{fig_fit}
\end{center}
\end{figure*}

Considering two (fast and slow) decay components of the scintillating crystals~\cite{PARK2002460,Kim:2014toa}, we characterize each pulse using a single rise time~($\tau_r$), two decay times ($\tau_f$ and $\tau_s$), the starting time of the pulse ($t_0$), and the ratio of the slow-to-fast decay components ($R$): 
\begin{equation}
		\begin{aligned}
				F_{i} (t) = & \frac{1}{\tau_{f_i}} e^{-(t-t_{0_i})/\tau_{f_i}} + \frac{R_i}{\tau_{s_i}} e^{-(t-t_{0_i})/\tau_{s_i}} \\
              & - ( \frac{1}{\tau_{f_i}} +\frac{R_i}{\tau_{s_i}} ) e^{(-t-t_{0_i})/\tau_{r_i}}, 
\end{aligned}
\label{func}
\end{equation}
Two-pulse events are modeled as $F(t) = A_1 F_1 (t) + A_2 F_2 (t)$ where $A_1$ and $A_2$ represent the total charge, proportional to the energy, of each pulse. 
Here, we can evaluate the energy of the first pulse (E1), the second pulse (E2), and $\Delta$T ($t_{0_2} - t_{0_1}$).
Mean decay times for two pulses calculated from the fit parameters are required to be greater than 150\,ns and less than 400\,ns for each pulse. 
Selected events are required to  have a reduced $\chi^2$ less than 4. 
In total, 2576 events are selected as the two-pulse events  from the 4258 candidate events. 
Figures~\ref{fig_fit} and \ref{fig_badtwoevents} show examples of the fit results  for the two-pulse candidate events that are categorized as two-pulse events and noisy events, respectively. 

For a quantitative measurement of the isomeric states, we need to evaluate selection efficiencies that depend on E1, E2, and $\Delta$T, and this requires simulated waveforms of the two-pulse events. 
Since photon simulations of the NaI(Tl) crystals are still being developed for the COSINE-100 experiment, we briefly report qualitative results based on our observations. 
Two-pulse events with small $\Delta$T, small E1 or E2, and large E1/E2 ratio  are preferentially eliminated by the initial selection of the candidate events. 
Furthermore, large E1/E2 ratio events are still contaminated by events with large tail fluctuations, even though applied selection criteria removed many of these events. 
Because of these limitations, we only study events with individual pulse energies in the range 2$-$30\,keV.

\begin{figure*}[!htb] 
\begin{center}
\begin{tabular}{cc}
\includegraphics[width=0.49\textwidth]{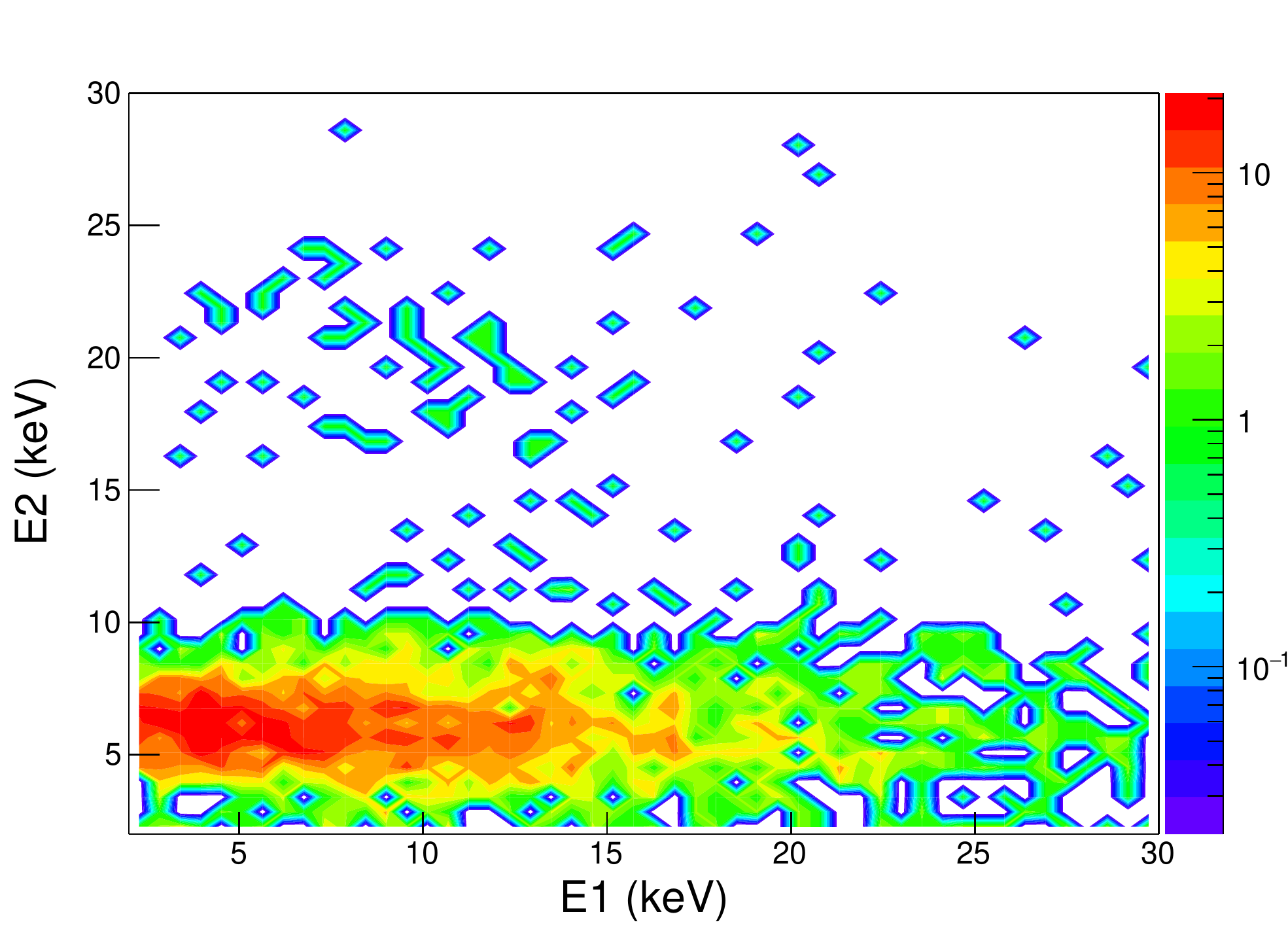}&
\includegraphics[width=0.49\textwidth]{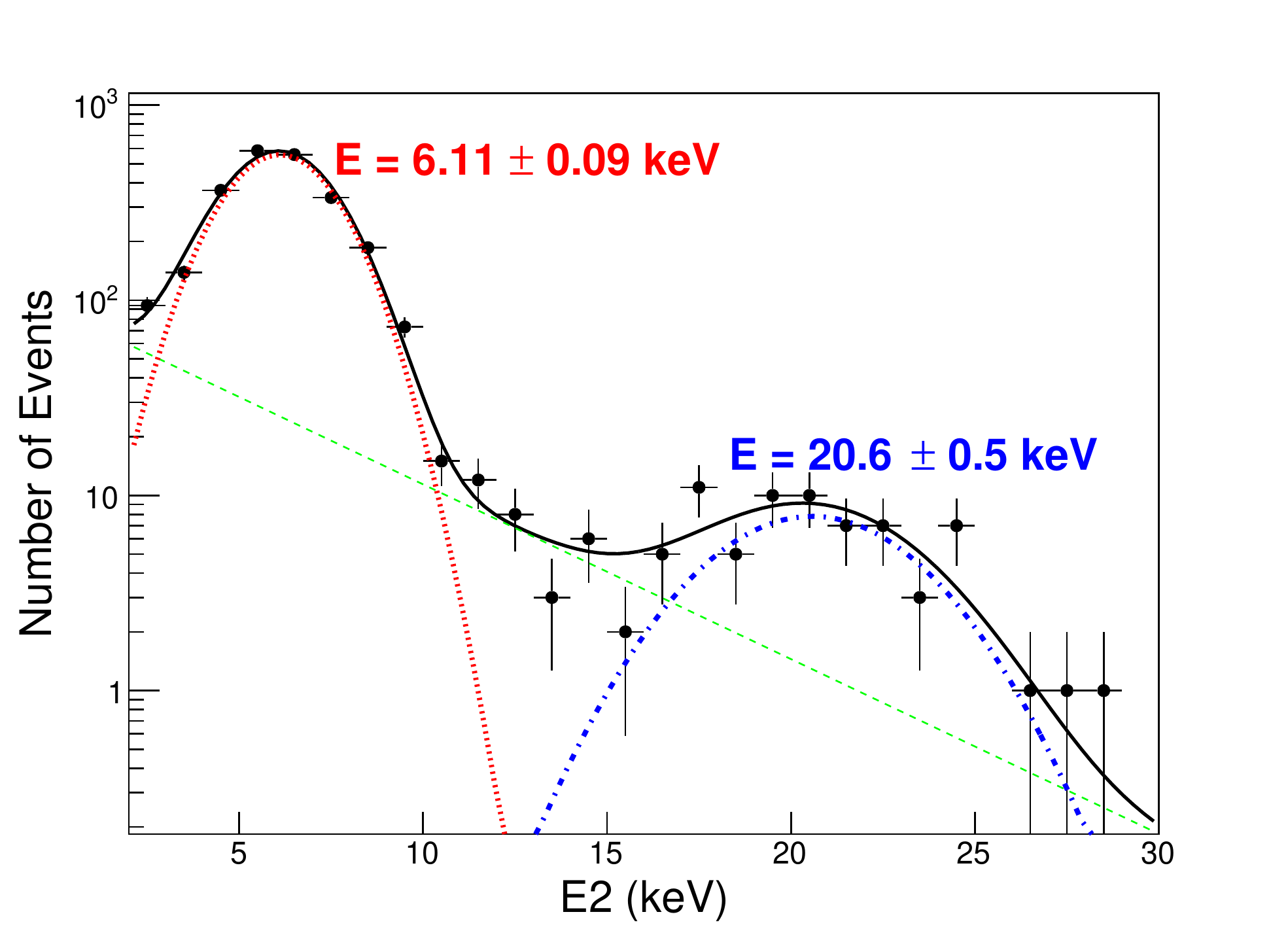}\\
(a) & (b) \\
\includegraphics[width=0.49\textwidth]{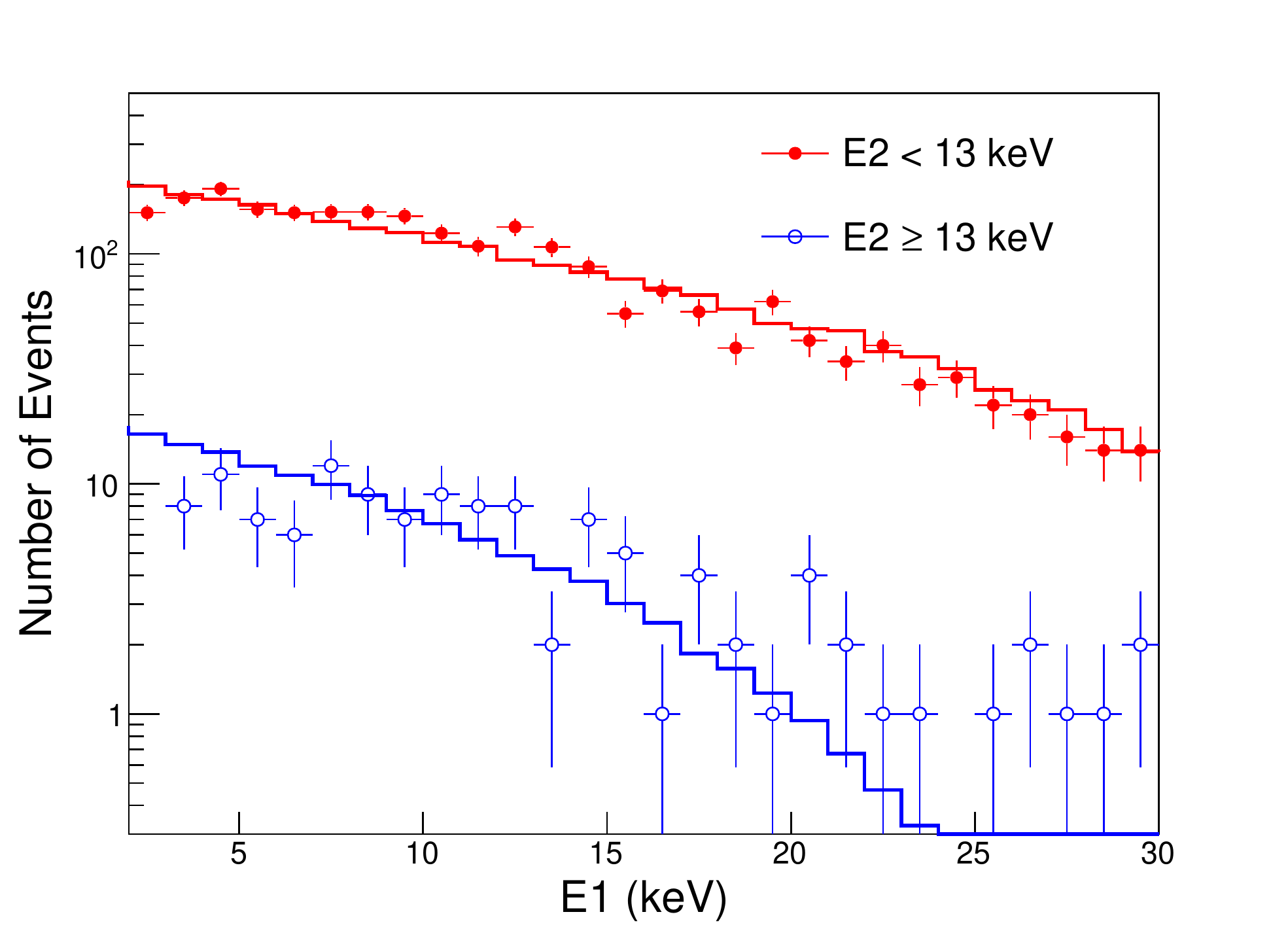}&
\includegraphics[width=0.49\textwidth]{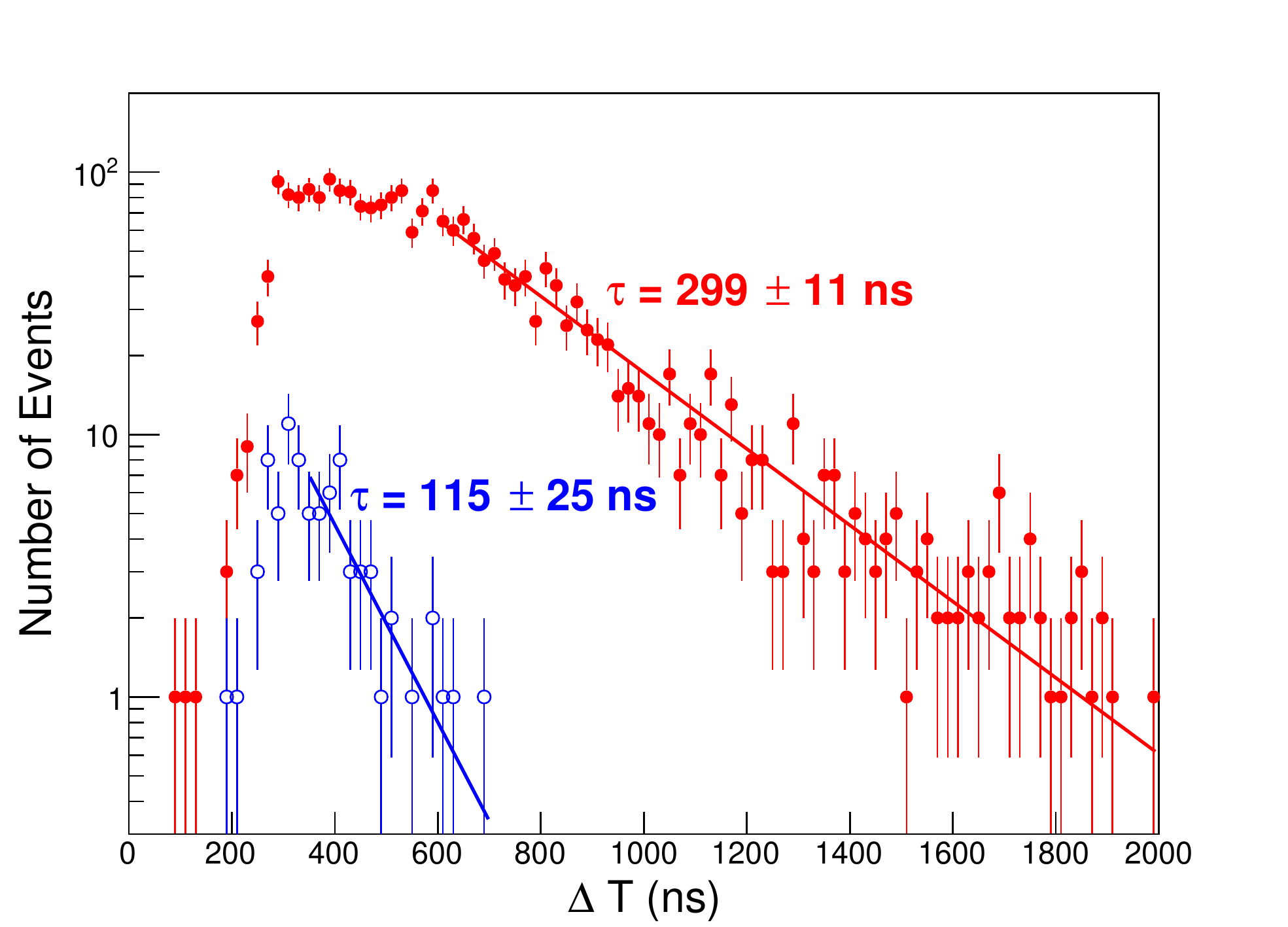}\\
(c) & (d) \\
\end{tabular}
\caption{(a) A scatter plot of energy of the second pulse {\it versus} the first pulse (E2 vs E1). (b) The E2 spectrum (points) is modeled with two Gaussians (dotted and dotted-dashed lines) and an exponential background contribution (dashed line). (c) The E1 spectra for E2$<$13\,keV (filled circles) and E2$\geq$13\,keV (opened circles) are compared with a Geant4-based simulation of $\beta$ spectra with Q=39.52\,keV (red solid line) and Q=25.61\,keV (blue solid line). (d) The distribution of the two-pulse time difference, $\Delta$T, for E2$<13$\,keV (red filled circles) and E2$\geq$13\,keV (blue opened circles). Exponential fits (solid lines) to obtain the lifetimes are overlaid.  
}
\label{fig_e1ve2}
\end{center}
\end{figure*}

Figure~\ref{fig_e1ve2} (a) shows a scatter plot of E1 and E2 for the 2--30\,keV energy region, 
where there is a large population of the two-pulse events around E2 equal to about 6\,keV.
The one dimensional spectrum of E2 is shown in Fig.~\ref{fig_e1ve2} (b), where there is a two-peak structure with peaks centered at around 6\,keV and 20\,keV. 
We model the E2 spectrum as two Gaussian functions summed with an exponential background. The exponential background describes the residual noisy events. 
The mean energies of the two peaks are determined to be: 6.11$\pm$0.09\,keV and 20.6$\pm$0.5\,keV, respectively, which closely
match the 6.28\,keV and 20.19\,keV nuclear levels of $^{228}$Ac, as indicated in Fig.~\ref{fig_Ra228}.

Figure~\ref{fig_e1ve2} (c) shows the E1 spectra for E2 values that are
greater or less than 13\,keV. The E2 spectra have energy distributions with shapes characteristic of mono-energetic emissions, while the E1 spectra have broader energy distributions that are characteristic of $\beta$-decay and match the decay scheme of $^{228}$Ra into $^{228}$Ac
shown in Fig.~\ref{fig_Ra228}.
The data are consistent with the identification of both the 6.28\,keV and 20.19\,keV
excited states of $^{228}$Ac as metastable isomers with lifetimes of $\mathcal{O}(100\,\mathrm{ns})$.
To confirm this hypothesis, we simulate the $\beta$ spectra of $^{228}$Ra using a Geant4-based simulation that is the same as the one used for background modeling of the COSINE-100 detectors~\cite{cosinebg,set2_background}. 
We model the energy spectra of two $\beta$-decays: one with Q = 39.52\,keV that decays into the 6.28\,keV state and the other with Q = 25.61\,keV that decays into the 20.19\,keV state. These energy spectra  are overlaid in Fig.~\ref{fig_e1ve2} (c) and show similar behaviors for energies greater than 5\,keV.  Some discrepancies can be explained at low energies due to the low efficiency of the small E1 events. 

Distributions of $\Delta$T are shown in Fig.~\ref{fig_e1ve2} (d) together with results from exponential fits for only large $\Delta$T events. 
This is because the two-pulse selection strongly suppress small $\Delta$T events. 
Only events with $\Delta$T$>$600\,ns are fitted for the E2$<$13\,keV distributions. 
Because two-pulse discrimination is more efficient for large E2 events, E2$>$13\,keV events with $\Delta$T$>$350\,ns are fitted. 
The fitted lifetimes are 299$\pm$11\,ns and 115$\pm$25\,ns for the E2 = 6.28\,keV and E2 = 20.19\,keV states, respectively, where  
only statistical uncertainties are considered.

\subsection{three-pulse events}
If we consider the level scheme of $^{228}$Ac as Fig.~\ref{fig_Ra228}, the 20.19\,keV state transits to the 6.67\,keV state before decaying to the ground state. Because of $\mathcal{O}(100\,\mathrm{ns})$ lifetime of  the 20.19\,keV state transition, the 20.19\,keV to 6.67\,keV transition must have the same lifetime. 
However, the number of two-pulse events with E2$>$13\,keV is only a few percent of those with E2$<$13\,keV. 
Considering the relative intensities of the $^{228}$Ra $\beta$-decay shown in Fig.~\ref{fig_Ra228}, the observed isomeric transition of the 20.19\,keV state is only $\mathcal{O}(1\%)$ of the total $\beta$-decay to the 20.19\,keV state. 
This may indicate that the 6.67\,keV state is also an isomeric state. In this case, three distinct emissions have to occur, but at a rate that cannot be seen in the current analysis. Only $\mathcal{O}(1\%)$ of these events would be accepted as the two-pulse events if the two final emissions occur so close in time that they cannot be distinguished in the NaI(Tl) crystal. 

\begin{figure*}[!htb] 
\begin{center}
		\begin{tabular}{ccc}
				\includegraphics[width=0.33\textwidth]{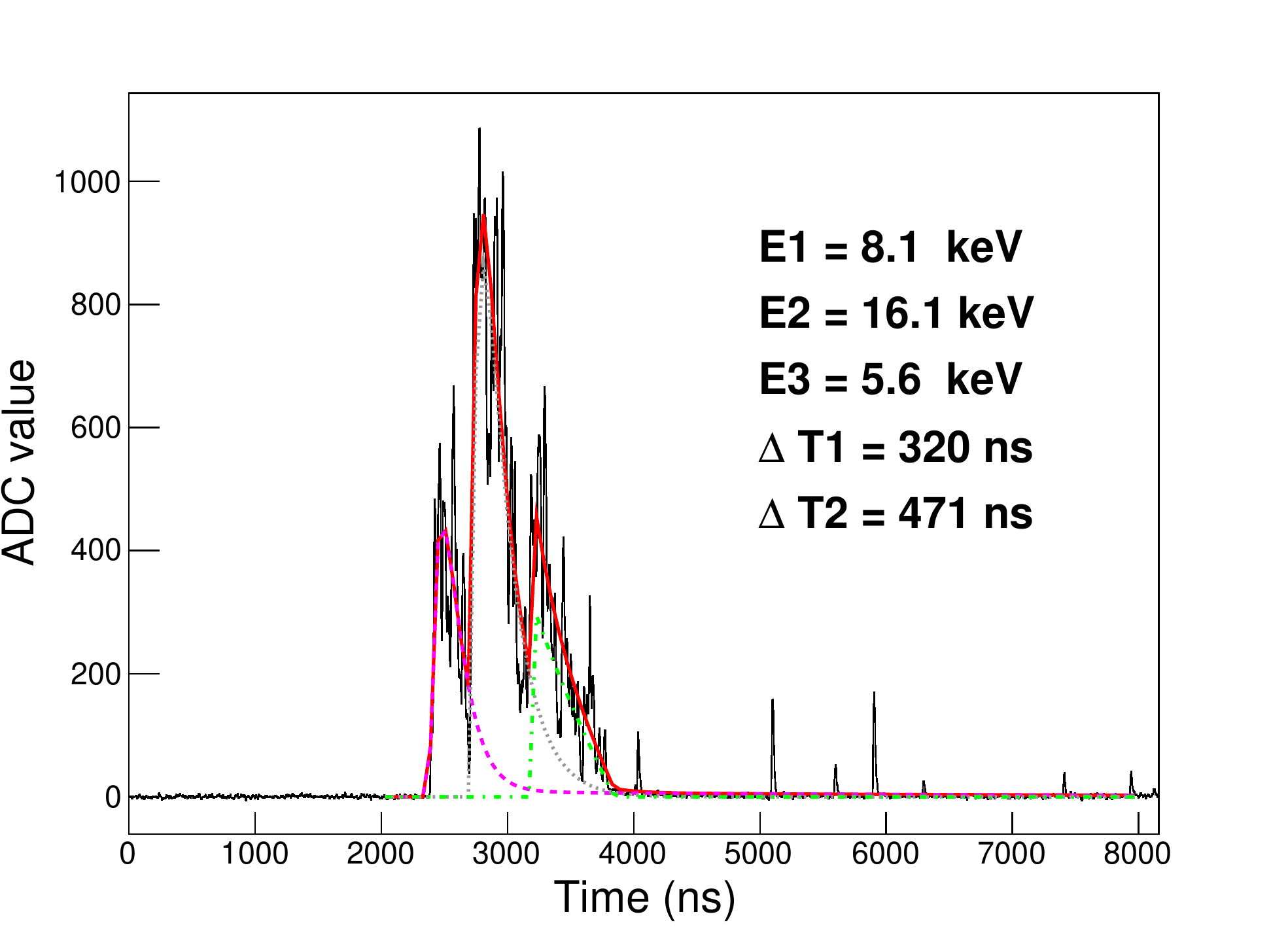} &
				\includegraphics[width=0.33\textwidth]{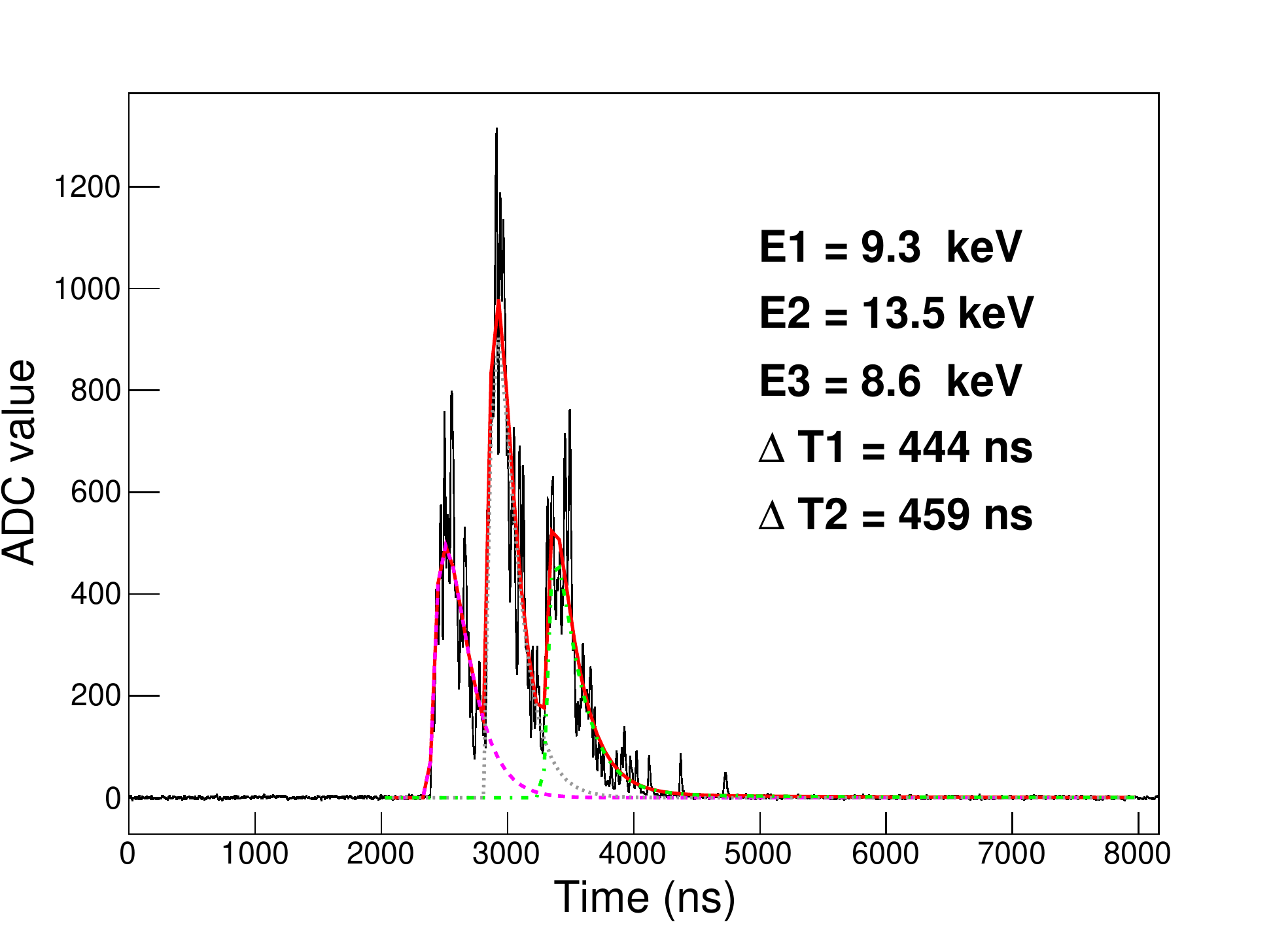} &
				\includegraphics[width=0.33\textwidth]{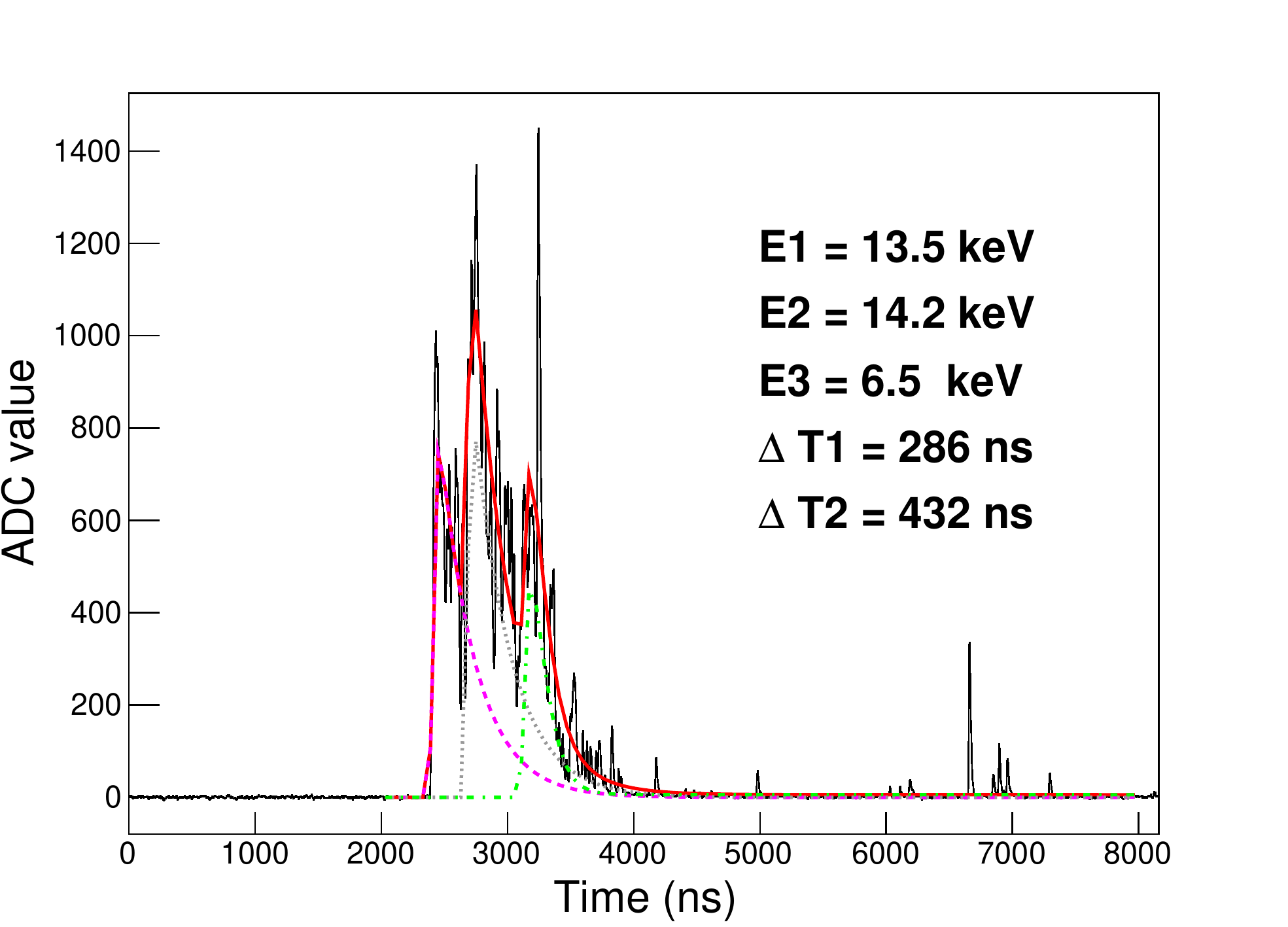} \\
Crystal 4 (Dec. 9, 2016) & Crystal 7 (July 12, 2018)& Crystal 3 (Aug. 20, 2019)\\
\end{tabular}
\caption{
Examples of the fit results for the three-pulse events from different crystals on different dates. The energies of each pulse (E1, E2, and E3) and the time differences ($\Delta$T1 and $\Delta$T2) are listed. 
}
\label{fig_threefit}
\end{center}
\end{figure*}

To idenfity the hypothesis of an isomeric 6.67\,keV state, we have searched for events that contain three distinct pulses, named ``three-pulse events'', within the 8\,$\mu$s window. 
Similar selections applied for the two-pulse events are used for the three-pulse events: $F(t) = A_1 F_1(t) + A_2 F_2(t) + A_3 F_3(t)$ from Eq.~\ref{func}. 
In total, 34 three-pulse events are selected.
Figure~\ref{fig_threefit} shows three examples of the fit results that occurred in three different crystals. 
Here, we evaluate the energy of the first pulse (E1), the second pulse (E2), and the third pulse (E3) and 
the time differences between first and second pulses ($\Delta$T1), and second and third pulses ($\Delta$T2). 
These quatities are indicated in the figure. 

\begin{figure*}[!htb] 
\begin{center}
\begin{tabular}{cc}
\includegraphics[width=0.49\textwidth]{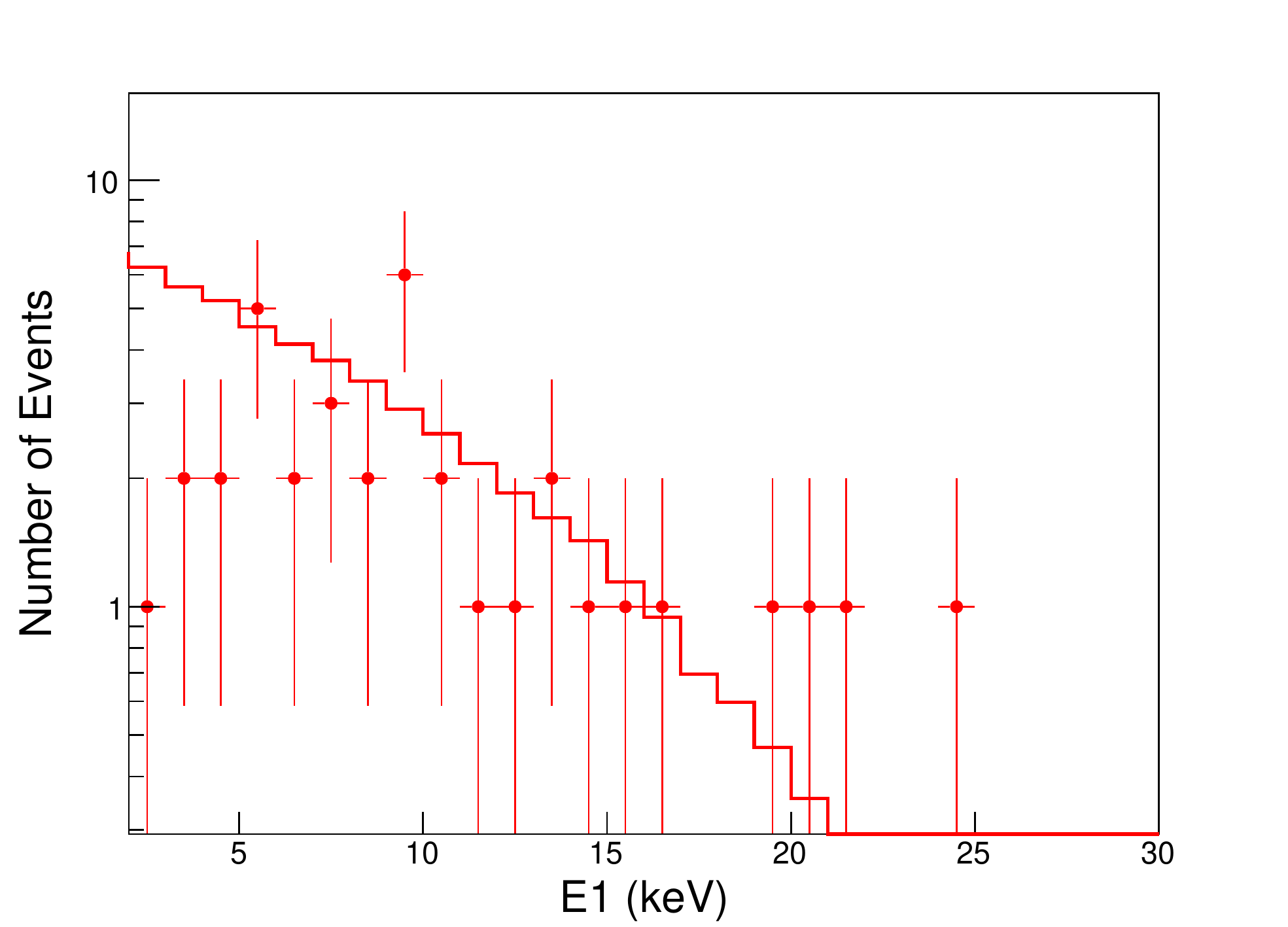}&
\includegraphics[width=0.49\textwidth]{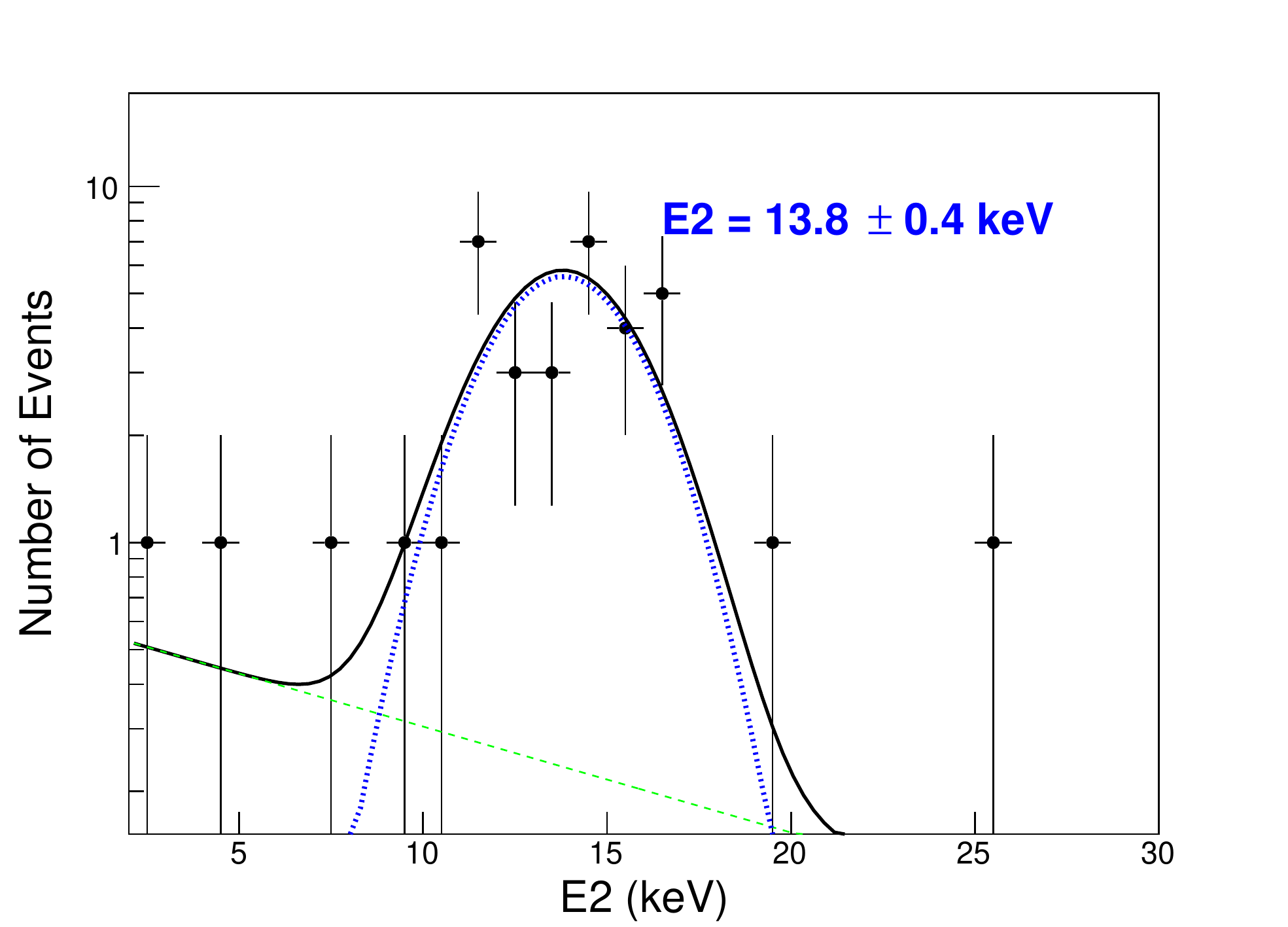}\\
(a) & (b) \\
\includegraphics[width=0.49\textwidth]{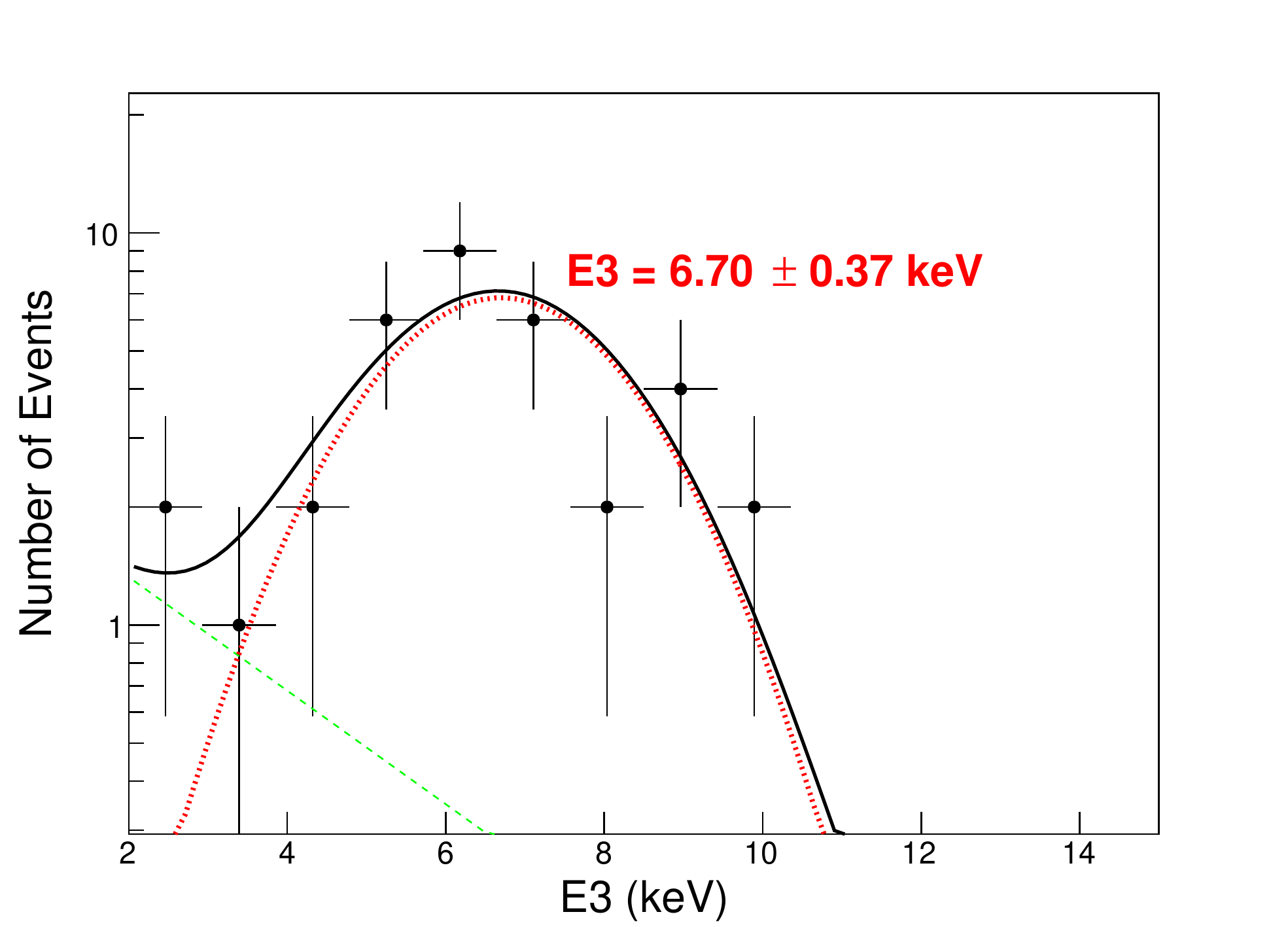}&
\includegraphics[width=0.49\textwidth]{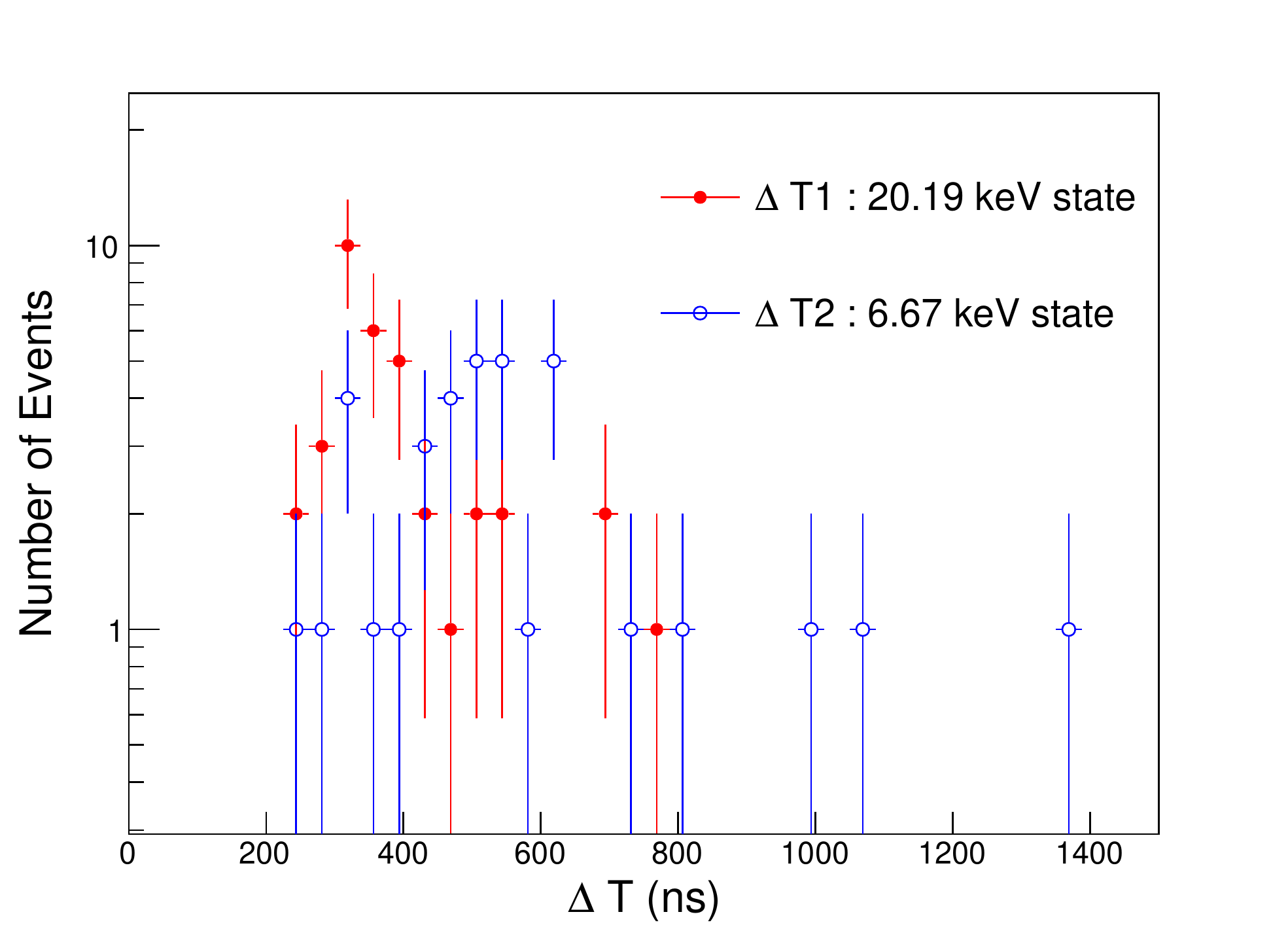}\\
(c) & (d) \\
\end{tabular}
\caption{ The energies and $\Delta$T distributions obtained from the three-pulse events are presented. (a) The E1 spectrum (points) of three-pulse events is compared with a Geant4-based simulation of $\beta$ spectra with Q=25.61\,keV (solid line). (b) The E2 spectrum (points) is modeled with a Gaussian (dotted line) and an exponential background contribution (dashed line). (c) The E3 spectrum (points) is modeled with a Gaussian (dotted line) and an exponential background contribution (dashed line). (d) The distributions of the two-pulse time differences from the first and second pulses, $\Delta$T1 (filled circle), and the second and third pulses, $\Delta$T2 (open circle), are presented. 
}
\label{fig_threeE}
\end{center}
\end{figure*}

Figure~\ref{fig_threeE} (a) shows the E1 spectrum that is overlaid with $\beta$ spectrum with Q = 25.61\,keV corresponding to decays into the 20.19\,keV state. Simulations and data show similar behavior for energies greater than 5\,keV. 
The energy spectra of the E2 and E3 are presented in Fig.~\ref{fig_threeE} (b) and (c), respectively. A Gaussian function with an exponential background fits the data well. 
The mean energies of E2 and E3 are determined as 13.8$\pm$0.4\,keV and 6.70$\pm$0.37\,keV, respectively. The obtained energy levels are well matched with $^{228}$Ac level scheme in Fig.~\ref{fig_Ra228}.
Figure~\ref{fig_threeE} (d) shows $\Delta$T1 and $\Delta$T2 distributions. We observe $\mathcal{O}(100\,\mathrm{ns})$ lifetime not only for the 20.19\,keV state but also for the 6.67\,keV state. 

\subsection{Consistency check of $^{228}$Ra decays}
As a cross-check of the $^{228}$Ac isomer hypothesis, we evaluate the level of $^{228}$Ra contamination in the NaI(Tl) crystals by measuring the rate of the two-pulse events with E2 around 6.28\,keV. 
We select 6.28\,keV state events by a requirement  of E2$<$13\,keV in Fig.~\ref{fig_e1ve2} (b). 
In order to account for the selection inefficiencies, we performed an extrapolation of the modeling in Fig.~\ref{fig_e1ve2} (c) and (d) for low-energy and small $\Delta$T events. 
Here, we assume a relative $\beta$ intensity of the 6.28\,keV state as 10\,\% according to data in Fig.~\ref{fig_Ra228}. 
Table~\ref{tab_activity} summarizes the measured $^{228}$Ra activities for the crystals that are compared with the results obtained from the standard background modeling of the COSINE-100 data~\cite{set2_background}. 
The consistency of the results for the $^{228}$Ra contamination supports the interpretation that the observed two-pulse events originate from $^{228}$Ra decays to isomeric excited states of $^{228}$Ac. 

\begin{table}[!ht]
\centering
\caption[Ra contamination]{$^{228}$Ra contamination in the COSINE-100 crystals measured using the two-pulse events at 6.28\,keV. These measurements are compared with the fit result from the background modeling of the COSINE-100 detector~\cite{set2_background}. }
\label{tab_activity}
\begin{tabular}{c|c|c}
\hline
(mBq/kg) & Two-pulses (this work) & Background modeling \\
\hline
C2 & 0.034 $\pm$ 0.010 & 0.032 $\pm$ 0.011 \\
C3 & 0.017 $\pm$ 0.005 & 0.029 $\pm$ 0.010 \\
C4 & 0.024 $\pm$ 0.007 & 0.012 $\pm$ 0.004 \\
C6 & 0.008 $\pm$ 0.002 & 0.024 $\pm$ 0.009 \\
C7 & 0.007 $\pm$ 0.002 & 0.015 $\pm$ 0.006 \\
\hline
\end{tabular}
\end{table}

\begin{figure}[!htb] 
\begin{center}
		\includegraphics[width=0.48\textwidth]{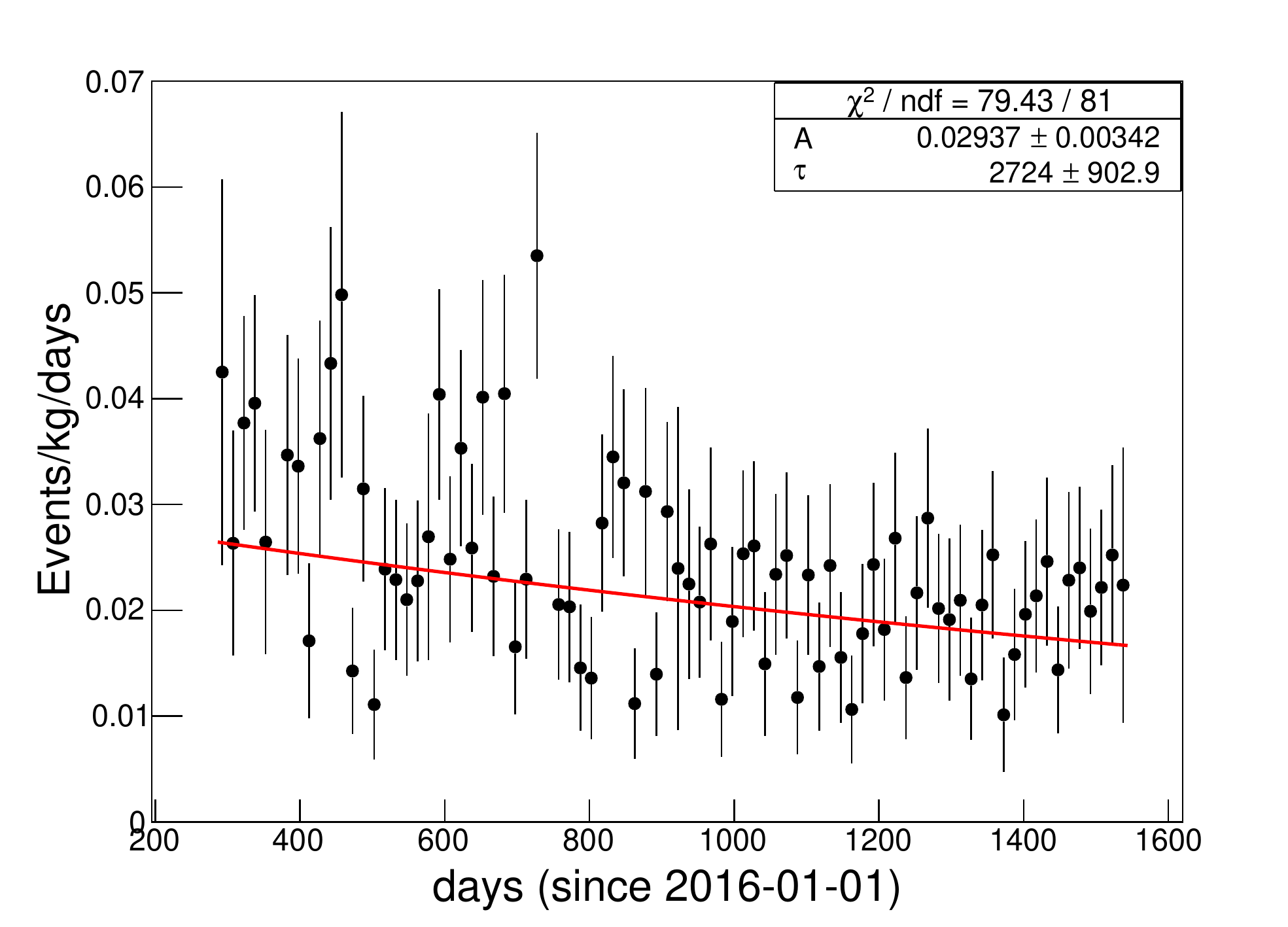} 
		\caption{
				 Time dependent event rates of the selected two-pulse events from the five crystals are modeled with an exponential decay component. 
}
\label{fig_trate}
\end{center}
\end{figure}

We have also evaluated the time dependent rate for the two-pulse events shown in Fig.~\ref{fig_trate}. A decreasing rate of  two-pulse events is evident and an exponential fit with $R(t)=A\exp(-t/\tau)$ is overlaid. Because initial purification of NaI powder~\cite{Shin:2018ioq,Shin:2020bdq} and decomposition of impurities from the crystal growing process~\cite{Park:2020fsq}, amounts of $^{232}$Th and $^{228}$Ra can be in non-equilibrium depending on their chemical properties. A similar non-equilibrium status between $^{238}$U and $^{226}$Ra was previously observed in the NaI(Tl) crystal~\cite{Park:2020fsq}. 
If crystalization effectively removes $^{232}$Th but not $^{228}$Ra, the initial $^{228}$Ra activity will be reduced with a lifetime of 3028\,days that is consistent with the measured rate decrease with $\tau$ = 2724$\pm$903\,days in Fig.~\ref{fig_trate}. This also supports  the interpretation that the observed two-pulse events originate from $^{228}$Ra decays. 

Based on the two- and three-pulse events measured in the COSINE-100 data, we conclude that the excited states of $^{228}$Ac at the 6.28\,keV, 6.67\,keV, and 20.19\,keV are isomers with lifetimes of $\mathcal{O}(100\,\mathrm{ns})$. 

\section{Impact on dark matter searches}
Internal contamination of dark matter detectors by $^{228}$Ra can introduce low-energy events by its $\beta$-decay and subsequent emission (total Q-value 45.8\,keV). 
Our observation shows that Q=39.52\,keV, 39.13\,keV, and 25.61\,keV $\beta$s and subsequent isomeric 6.28\,keV, 6.67\,keV, and 20.19\,keV emissions occur with  $\mathcal{O}(100\,\mathrm{ns})$ time differences. 
Depending on the specific  characteristics and data analysis methods of an experiment, the $\beta$ and the following emissions from $^{228}$Ac could be identified as two separate events due to their $\mathcal{O}(100\,\mathrm{ns})$ time separation.
In the case of the COSINE-100 detector, we measure the energy deposited in our detectors by integrating the signal over a 5\,$\mu$s time window, resulting in the $\beta$ particle and the following emission being treated  as a single event~\cite{Adhikari:2018ljm,Adhikari:2019off}. 
Therefore, the background modeling using the existing Geant4-based simulation is sufficient for our current analysis~\cite{cosinebg,set2_background}.
However, if one uses fast response detectors, such as organic scintillators that have a less than 50\,ns decay time~\cite{SJOLIN196545}, the coincident $\beta$ and $\gamma$ or conversion electron with  $\mathcal{O}(100\,\mathrm{ns})$  time difference could be identified as separate events. 
In the extreme case, this could result in only the first pulses being properly accounted for while the delayed pulses would be ignored because of being too close to a previous event.

\begin{figure}[!htb] 
\begin{center}
		\includegraphics[width=0.48\textwidth]{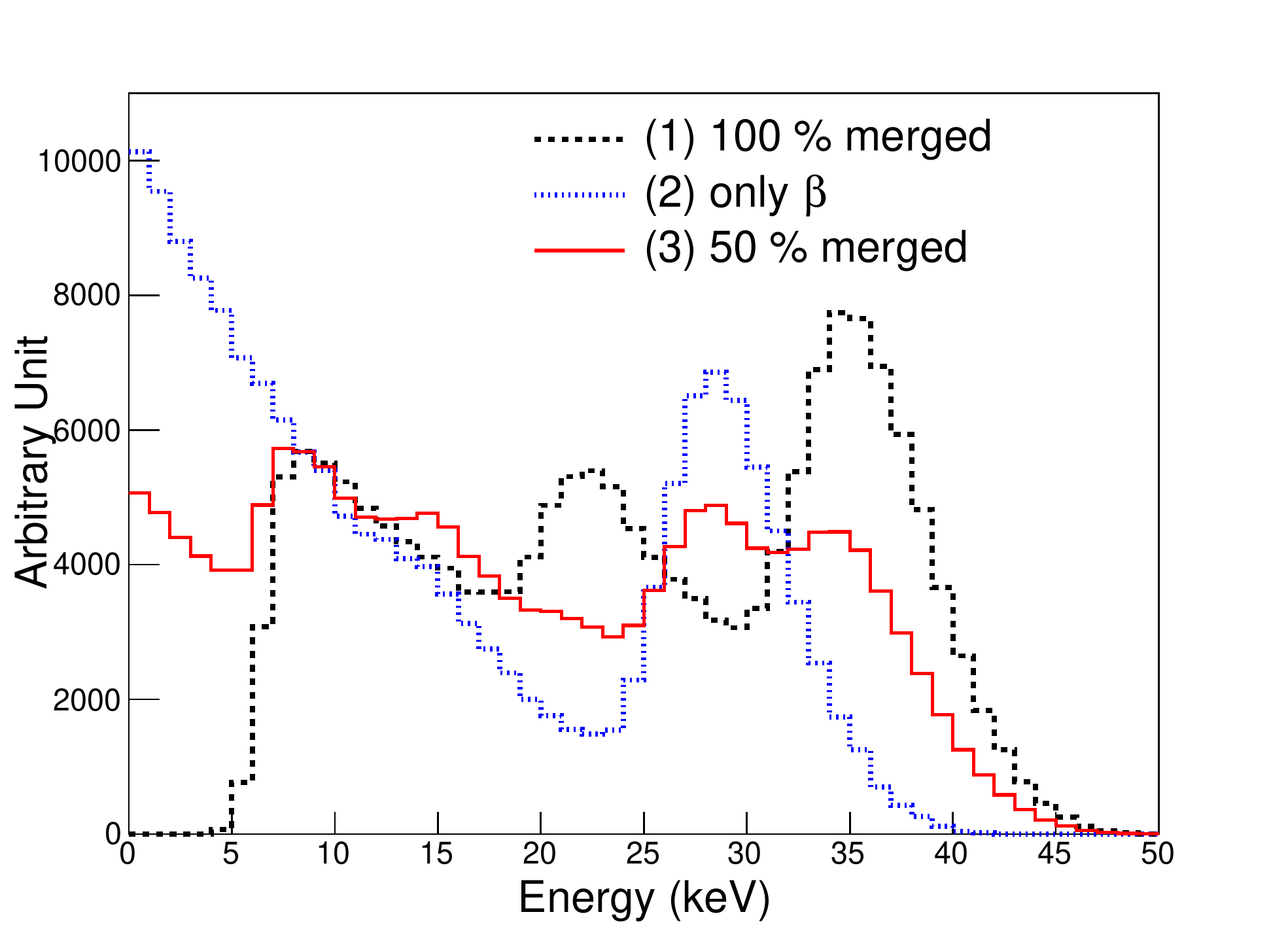} 
		\caption{
  $^{228}$Ra decay spectra from Geant4-based simulation assuming a few different cases of experiments measuring $^{228}$Ac isomeric decays. Case (1) considers that all isomeric decays in the three excited states are merged with the initial electrons from the $\beta$-decay. Case (2) considers that only the initial electrons are accounted. Case (3) is half of the case (1) and (2). 
}
\label{fig_Ra228spec}
\end{center}
\end{figure}

We have estimated the impact of these isomeric states of $^{228}$Ac using a Geant4-based simulation, with the result shown in Fig.~\ref{fig_Ra228spec} for three different cases: (1) all isomeric decays are merged with electrons in the $\beta$-decay, (2) only the $\beta$-decay electrons for the three isomeric states are observed, and (3) 50\% events are merged and the other accounts only the initial electrons. 
Here, we use the energy resolution of the COSINE-100 detector and assume a 100\,\% detection efficiency for our convenience. 
In case (1), electron energies from the $\beta$ emissions are merged with the following mono-energetic $\gamma$ or conversion electron emissions (6.28\,keV, 6.67\,keV, 20.19\,keV and 33.1\,keV) and 
there are almost no events below 6\,keV that can be seen in Fig.~\ref{fig_Ra228spec}. 
If the isomeric emissions are perfectly distinguished with the initial electrons from $\beta$-decay and the following isomeric emissions are not used  as like the case (2), only $\beta$ spectra from decays into the three isomeric states are shown. In this case, large populations in the low energy signal region are presented as one can see in Fig.~\ref{fig_Ra228spec}. Case (3) is mixture of the cases (1) and (2) but, it also accounts for 13.52\,keV isomeric emission summed with the initial electrons while the isomeric 6.67\,keV decays are distinguished. 
The mixture of cases (1), (2), and (3) would be different for each experiment depending on the detector performance and the analysis technique. 
Therefore, the impact of these isomer states should be studied by each dark matter search experiment.

\section{Conclusion}
In conclusion, we have identify the isomers at 6.28\,keV, 6.67\,keV, and 20.19\,keV states in $^{228}$Ac from the $\beta$-decay of $^{228}$Ra with the COSINE-100 detector. 
Their lifetimes are measured to be 299$\pm$11\,ns and 115$\pm$25\,ns for 6.28\,keV and 20.19\,keV, respectively, although only statistical uncertainties are obtained. 
Due to the low Q-values of these $\beta$-decays, these isomeric states have the potential to  impact the background modeling of the dark matter search experiments in the low-energy signal region. 
Thus, a dedicated experiment to measure the nuclear structure of $^{228}$Ac is highly desirable.

\begin{acknowledgements}
We thank the Korea Hydro and Nuclear Power (KHNP) Company for providing underground laboratory space at Yangyang.
This work is supported by:  the Institute for Basic Science (IBS) under project code IBS-R016-A1 and NRF-2016R1A2B3008343, Republic of Korea;
NSF Grants No. PHY-1913742, DGE-1122492, WIPAC, the Wisconsin Alumni Research Foundation, United States; 
STFC Grant ST/N000277/1 and ST/K001337/1, United Kingdom;
Grant No. 2017/02952-0 FAPESP, CAPES Finance Code 001, CNPq 131152/2020-3, Brazil.
\end{acknowledgements}

\providecommand{\href}[2]{#2}\begingroup\raggedright\endgroup
\end{document}